\documentstyle[12pt,amsfonts,epsfig]{article}                  

\makeatletter
\def\nothing#1{}
\newdimen\earraycolsep
\renewcommand{\theequation}{\arabic{section}.\arabic{equation}}
\renewcommand{\thesection}{\arabic{section}}

\renewcommand{\thetable}{\arabic{table}}
\renewcommand{\thefigure}{\arabic{figure}}
\def\author#1{{\pretolerance=10000 \raggedright \advance \leftskip by 1in \noindent #1 \vskip 1pc}}
\def\affiliation#1{{\advance\leftskip by 1in \noindent #1 \vskip -1pc}}

\renewcommand\section{\@startsection{section}{1}{\z@}{2pc \@plus 
      1ex minus .2ex}{1pc \@plus .2ex}{\reset@font
      \normalsize\bfseries\noindent }}
\def\abstract{\section*{ABSTRACT}}
\renewcommand\subsection{\@startsection{subsection}{2}{\z@}{1pc \@plus 1ex
    minus.2ex}{1pc \@plus .2ex}
    {\reset@font\normalsize\bfseries }}
\renewcommand\subsubsection{\@startsection{subsubsection}{3}{\parindent}
        {1pc \@plus 1ex minus.2ex}{-0.5em}{\reset@font\normalsize\bfseries}}  

\def\AmS{{\protect\the\textfont2%
        A\kern-.1667em\lower.5ex\hbox{M}\kern-.125emS}}

\def\p@LaTeX{{\family{times}\series{m}\shape{n}\selectfont L\kern-.36em\raise.3ex\hbox{\scriptsize A}\kern-.15em T\kern-.1667em\lower.7ex\hbox{E}\kern-.125emX}}
\renewcommand{\thefootnote}{\fnsymbol{footnote}}
\newlength{\colwidth}

\def\@oddhead{\hfil}
\def\@evenhead{\hfil}
\def\@oddfoot{{\bfseries\hfil\thepage}}
\def\@evenfoot{{\bfseries\thepage\hfil}}
\def\fnum@figure{\footnotesize\raggedright{\bfseries \figurename~\thefigure.}}
\def\fnum@table{\normalsize\raggedright{\bfseries \tablename~\thetable.}}
\long\def\@makecaption#1#2{\vskip 10\p@ {#1 #2\par}}
\long\def\@makefntext#1{\setbox0=\hbox{$\m@th^{\@thefnmark}$}\noindent\hangindent=\wd0 \box0 #1}
\def\centerfig#1#2#3#4{\vspace*{#2}\relax\centerline{\hbox to#1{\special{#4:#3.#4 x=#1, y=#2}\hfil}}}
\newbox\@atbox
\long\def\atable#1#2#3{\begin{table}[tbp]\centering\footnotesize
\setbox\@atbox\hbox{#2}
\parbox{\wd\@atbox}{\caption{#1}}\par\smallskip
#2
\par\smallskip\parbox{\wd\@atbox}{\raggedright #3}
\end{table}}
\def\@nbibitem#1{\noindent \hangindent=2pc \hangafter=1
\refstepcounter{enumi}\hbox to 2pc{\arabic{enumi}.\hfil}%
\immediate\write\@auxout{\string\bibcite{#1}{\arabic{enumi}}}}
\def\numbibliography{%
\section*{REFERENCES}%
\bgroup\footnotesize
\setcounter{enumi}{0}%
\def\newblock{\hskip .11em plus.33em minus.07em}%
\let\bibitem\@nbibitem}
\def\endnumbibliography{\par\egroup}
\makeatother

\def\nn{\nonumber}
\def\non{\nonumber\\}
\def\be{\begin{equation}}
\def\ee{\end{equation}}
\def\ben{\begin{displaymath}}
\def\een{\end{displaymath}}
\def\ba{\begin{eqnarray}}
\def\ea{\end{eqnarray}}

\def\a{\alpha}
\def\b{\beta}
\def\D{\Delta}
\def\d{\delta}
\def\e{\varepsilon}
\def\f{\varphi}

\def\g{\gamma}

\def\l{\lambda}

\def\m{\mu}
\def\n{\eta}
\def\o{\omega}
\def\O{\Omega}

\def\r{\rho}
\def\s{\sigma}

\def\t{\tau}

\def\x{\xi}

\def\e{\epsilon}
\def\tP{{\tilde \Phi}}

\def\cC{{\cal C}}
\def\E{{\cal E}}
\def\H{{\cal H}}
\def\cL{{\cal L}}
\def\cO{{\cal O}}
\def\cF{{\cal F}}

\def\cD{{\cal D}}
\def\cV{{\cal V}}

\def\cM{{\cal M}}
\def\cH{{\cal H}}
\def\cP{{\cal P}}
\def\E{{\cal E}}
\def\B{{\cal B}}
\def\eb{{\bf e}}
\def\fb{{\bf f}}
\def\hb{{\bf h}}

\def\gb{{\bar \gamma}}

\def\sh{{\hat{\s}}}
\def\vh{{\hat{\cV}}}

\def\Pt{\tilde{P}}

\def\Bs{B}

\def\C{{\mathbb{C}}}

\def\R{{\mathbb{R}}}

\def\la{\label}
\def\ci{\cite}

\def\Ref#1{(\ref{#1})}

\def\f{\frac}
\def\ft#1#2{{\textstyle {\frac{#1}{#2}} }}
\def\i{\infty}
\def\p{\partial}

\def\tr{{\rm tr}}
\def\ra{\rightarrow}

\def\coset{SL(2,\R)/SO(2)}

\def\rt{{\tilde \rho}}
\def\et{{\tilde \varepsilon}}
\def\gt{{\tilde h}}
\def\pa{\partial}
\def\Pit{{\tilde \Pi}}
\def\Pt{{\tilde P}}
\def\Ph{{\widehat P}}
\def\Rh{{\widehat R}}
\def\dxy{{\d (x-y)}}

\begin{document}
\begin{flushright}
DESY 96-249\\
hep-th/9612065\\
December 1996\\
\vspace*{1cm}
\end{flushright}
\begin{center}
{\large \bf INTEGRABLE CLASSICAL AND QUANTUM GRAVITY\footnote{Lectures
given by H.~Nicolai at NATO Advanced Study Institute on Quantum Fields
and Quantum Space Time, Carg\`ese, France, 22 July - 3 August.} }
\bigskip\medskip\\
{\bf H.~Nicolai, D.~Korotkin,\footnote{On leave of absence from
Steklov Mathematical Institute, Fontanka, 27, St.Petersburg 191011
Russia} H.~Samtleben}\bigskip\\
II. Institut f\"ur Theoretische Physik \\
Universit\"at Hamburg\\Luruper Chaussee 149\\22761 Hamburg, Germany\\
{\small E-Mail: nicolai@x4u2.desy.de, korotkin@x4u2.desy.de, 
jahsamt@x4u2.desy.de}
\end{center}
\renewcommand{\thefootnote}{\arabic{footnote}}
\setcounter{footnote}{0}
\vspace*{1.2cm}


\section{INTRODUCTION}     
In these lectures we report recent work on the exact quantization of
dimensionally reduced gravity
\ci{KorNic95a,KorNic95b,KorNic96,KorSam96}. Assuming the presence
of commuting Killing symmetries which effectively eliminate the 
dependence on all but two space-time coordinates allows us to cast the 
models into the form of $2d$ non-linear ($G/H$)-coset space $\s$-models 
coupled to gravity and a dilaton. This construction includes a variety
of models described by different coset spaces --- ranging from pure
$4d$ Einstein gravity with two commuting Killing vector fields 
to dimensionally reduced maximal supergravity with the coset 
space $E_{8(+8)}/SO(16)$ \ci{BJ1}. Although these models 
superficially  resemble the $2d$ dilaton-gravity models considered 
more recently in the context of string theory and Liouville theory 
(see \ci{dila} and references therein),  the latter generically admit
more general couplings of the dilaton and the Liouville field while
restricting the matter sector essentially to a set of free fields.
By contrast, the matter sector of the models considered here is
governed by highly non-linear interactions; its underlying group
theoretical structure is a crucial ingredient in our analysis. 

In terms of the unified formulation presented in these lectures the
models allow consistent quantization by use of methods developed over
many years in the context of flat space integrable systems \ci{Fadd84,
FadTak87}. More specifically, we will show that the Wheeler-DeWitt
(WDW) equation can be reduced to a modified version of the
Knizhnik-Zamolodchikov (KZ) equations from conformal theory
\ci{KniZam84}, the insertions being given by singularities in the spectral
parameter plane. This basic result in principle permits the explicit
construction of solutions, i.e.~physical states of the quantized
theory. In this way, we arrive at integrable models of quantum gravity
with infinitely many self-interacting propagating degrees of
freedom. We here would like to emphasize not just the technical
aspects of the construction, but also the fact that our results may
serve to investigate various conceptual issues of quantum gravity (see
\ci{Hawk84,Ish91,Asht91} for introductory reviews with many further
references).

The lectures are divided into four parts. We first describe how the
dimensional reduction of $3d$ gravity-coupled non-linear $\s$-models
gives rise to the particular $2d$ models studied in these lectures, of
which pure Einstein gravity with two commuting Killing vectors is the
simplest example. Chapter 3 deals with the classical integrability of
the model, that is manifest in the existence of a linear system. The
spectral parameter current is introduced as new fundamental quantity
and the dynamics studied in these terms. Restriction to an
isomonodromic ansatz results in a set of decoupled differential
equations completely describing the dynamics; the relaxation of these
restrictions is briefly sketched in the end. In Chapter 4 we set up a
Hamiltonian framework for the model, such that translations along the
light-cone are governed by two independent Hamiltonians and the phase
space is essentially attached to one point in space-time. We compare
the new scheme to the conventional Hamiltonian approach and discuss
some open questions.  Exploiting the new Hamiltonian structure, we
develop the quantization of the model in Chapter 5. The Wheeler-DeWitt
equations that identify physical states in the quantum theory are
explicitly stated. Physical states solving these equations are built
from solutions of a modified version of the Knizhnik-Zamolodchikov
system. Finally, the appendix sketches the extension of the formalism
to supersymmetric models.

\section{NONLINEAR $\s$-MODELS COUPLED TO GRAVITY}

After dimensional reduction, the gravity and supergravity models
mentioned in the introduction lead to $3d$ gravity-coupled coset
$\s$-model with various coset spaces $G/H$ (see \ci{BrMaGi88} for a
systematic discussion). We start by describing these models and
further reducing them to two dimensions.

\subsection{Nonlinear $\s$-Models in Three Dimensions}

The models we are going to consider in these lectures are most
conveniently obtained by dimensional reduction of the following
non-linear $\s$-model in three dimensions coupled to gravity
\be\la{Lag3d}
\cL = -\ft12 eR(e)+\ft12 eh^{mn}~\tr P_mP_n.
\ee
The first term is the usual Einstein-Hilbert action for the $3d$
metric $h_{mn}=e_m^{~a}e_n^{~b}\n_{ab}$ with $e\equiv\det e_m^{~a}$
and $\n_{ab}$ the flat (Minkowski) metric, where indices $m, n,\dots$
and $a, b, \dots$ label curved and flat space vectors in three
dimensions.  The matter sector of this model, governed by the second
term in \Ref{Lag3d}, is based on a set of scalar fields, which are
combined into a matrix $\cV(x)$ taking values in a non-compact Lie
group $G$ with the maximal compact subgroup $H$.  This subgroup can be
characterized by means of a symmetric space involution
$\n:G\rightarrow G$ as
\ben
H=\{{\rm h}\!\in\!G~|~\n({\rm h})\!=\!{\rm h}\}
\een
The involution extends naturally to the Lie algebras ${\mathfrak
g}\!=\!{\rm Lie}~G$ and ${\mathfrak h}\!=\!{\rm Lie}~H$, respectively,
such that ${\mathfrak g}\!\equiv\!{\mathfrak h}\oplus{\mathfrak k}$ is
a direct sum of the eigenspaces of $\n$, orthogonal with respect to
the Cartan-Killing form. The maximality of the coset space is
expressed by 
\ben
[{\mathfrak h},{\mathfrak h}] \subset {\mathfrak h},\qquad
[{\mathfrak h},{\mathfrak k}] \subset {\mathfrak k},\qquad
[{\mathfrak k},{\mathfrak k}] \subset {\mathfrak h}
\een
In terms of a basis $Z^a$ of $\mathfrak g$ 
($a,b,...=1,..., {\rm dim} \, {\mathfrak g}$)\footnote{Since flat
$3d$ Minkowski indices will not play any role after this chapter,
we hope that this double usage of indices will not cause undue confusion.}, 
the commutation relations are given by
\be
[Z^a,Z^b] = {f^{ab}}_c Z^c
\ee
Whenever necessary we will distinguish the subgroup and coset generators 
by writing $Z^\a$ and $Z^A$, respectively, where $\a,\b,...=1,...,
{\rm dim}\, {\mathfrak h}$ and $A,B,...$ run over the remaining indices.
As an example take
$G\!=\!SL(n,\R)$: its maximal compact subgroup $H\!=\!SO(n)$ 
is characterized by the involution $\n({\rm g})\!\equiv\!({\rm
g}^t)^{-1}$ for ${\rm g}\!\in\!G$ or $\n(X)\!\equiv\!-X^t$ for
$X\!\in\!\mathfrak{g}$, respectively. 

The decomposition
\be\la{PQ}
\cV^{-1}\p_m\cV ~\equiv~ Q_m + P_m ~\in~ {\mathfrak g} 
\ee
defines the quantities $Q_m\!\equiv\!Q_m^\a Z_\a\in{\mathfrak h}$
and $P_m\!\equiv\!P_m^AZ_A\in{\mathfrak k}$ appearing in the
Lagrangian \Ref{Lag3d}. Due to their definition they are subject to
the compatibility relations (valid in any dimension)
\ba
\p_m Q_n - \p_nQ_m + [Q_m,Q_n] &=& -[P_m,P_n] \la{corel}\\
D_mP_n - D_nP_m &=& 0 \nn
\ea
with $D_mP_n\equiv \p_mP_n + [Q_m,P_n]$. 
\bigskip

The Lagrangian \Ref{Lag3d} is invariant under the transformations
\be\la{gauge}
\cV(x)\mapsto {\rm g}^{-1}\cV(x){\rm h}(x)
\ee
where ${\rm g}\!\in\!G$ is constant and ${\rm h}(x)\!\in\!H$ is 
a local (gauge) transformation. Consequently, the composite fields
$P_m$ and $Q_m$ are inert w.r.t.~to the rigid $G$ invariance, but
do transform under $H$ according to 
\be
P_m\!\mapsto\!{\rm h}^{-1}P_m{\rm h} \;\;\; , \;\;\;
Q_m\!\mapsto\!{\rm h}^{-1}Q_m{\rm h}\!+\!{\rm h}^{-1}\p_m{\rm h}
\ee
The transformations \Ref{gauge} are analogous to the
transformation properties of the vierbein (tetrad) in
general relativity, where the rigid transformations represent the
freedom of constant linear coordinate transformations whereas the
local Lorentz transformations correspond to the invariance of the
metric under different decompositions into the vielbein. The latter
(local) freedom may be used to parametrize the physical fields 
and the coset space $G/H$ by a fixed system of representative
elements of $G$. E.g., for $SL(n,\R)/SO(n)$
it is often convenient to choose a triangular form of the matrices.

The equations of motion derived from \Ref{Lag3d} read
\be
D_m(eh^{mn}P_n) = 0
\ee
together with Einstein's equations for the metric $h_{mn}$.
\bigskip

As the simplest example for these models, let us briefly recall how
the dimensional reduction of pure Einstein gravity from four dimensions
to three dimensions leads to a Lagrangian of the type \Ref{Lag3d}. The
original vierbein may be brought into triangular form by means of
a local Lorentz transformation:
\be\la{4bein}
E_M^{~A}=
\left(\begin{array}{cc} \D^{-\frac12}e_m^{~a} & \D^{\frac12}B_m\\
        0 & \D^{\frac12}\end{array} \right)
\ee
With this parametrization and dropping the dependence on the fourth
(spatial) coordinate, the equations of motion can be equivalently
obtained from the Lagrangian:
\be\la{Lag43}
\cL^{(3)} = -\ft12eR^{(3)}(e) + 
\ft14eh^{mn}\D^{-2}(\p_m\D\p_n\D+\p_mB\p_nB)~,
\ee 
where $B$ is dual to the Kaluza-Klein vector $B_m$:
\ben
\D^2(\p_mB_n-\p_nB_m) \equiv \e_{mnp}\p^p B,
\een
The scalar fields $\D$ and $B$ descending from the vierbein
\Ref{4bein} now build the matter part of the $3d$ model \Ref{Lag43},
and are coupled to $3d$ gravity (which carries no propagating degrees
of freedom any more). They correspond to the two helicity states of
the graviton, with $SO(2)$ as the helicity subgroup of $SO(1,3)$. The
Lagrangian \Ref{Lag43} is a nonlinear $\s$-model of the type
\Ref{Lag3d} with the triangular matrices
\be\la{Vpar}
\cV = \left(\begin{array}{cc} \D^{\frac12} & \D^{-\frac12}B\\
0 & \D^{-\frac12} \end{array} \right) \in SL(2,\R)
\ee
which build a representative system of the coset space $\coset$. 

\subsection{Reduction to Two Dimensions}
The reduction of the $3d$ model to two dimensions is achieved dropping
the dependence on another coordinate. Depending on the norm of the
corresponding Killing vector, the $2d$ model will live on an Euclidean
or Lorentzian worldsheet, respectively. While the former reduction of
Einstein's theory corresponds to stationary axisymmetric solutions,
the latter can describe physically inequivalent solutions, namely (in
the free field truncation) Einstein-Rosen gravitational waves
\ci{EinRos37}, or colliding plane waves \ci{KhanPen71}. In these
lectures we will concentrate on the second case, i.e.~Lorentzian
signature worldsheets.  However, in a very formal sense, the two cases
are related by a Wick rotation from real to imaginary time.

Consider again a triangular form of the dreibein $e_m^{~a}$
which we parametrize as
\be\la{3bein}
e_m^{~a} = \left(\begin{array}{cc} e_\m^{~\a} & \r A_\m\\0&\r
               \end{array} \right)
\ee
where $e_\m^{~\a}$ is the zweibein (dyad), and
Greek indices label the remaining two space-time dimensions.
{}By the field equations, the Kaluza-Klein vector field $A_\m$
carries no dynamical degrees of freedom; assuming absence of a 
cosmological constant we can thus ignore it.
The reduced Lagrangian is given by
\be\la{Lag42}
\cL^{(2)} = -\ft12 \r eR^{(2)}(e) + \ft12 \r eh^{\m\nu}\tr P_\m P_\nu
\ee
where $e$ is now the zweibein determinant. The appearance of the
dilaton field $\r$ is a typical feature of Kaluza-Klein type
dimensional reduction. Namely, this field ``measures" the size of the
compactified dimensions of the higher-dimensional space-time: for the
direct reduction of the vielbein $E_M^{~A}$ from $d$ to two dimensions
(as opposed to the detour via three dimensions we are taking here)
\ben
E_M^{~A} = \left(\begin{array}{cc} e_\m^{~\a} & * \\
                  0& E_{M'}^{~A'} \end{array} \right)
\een
where $M'$ and $A'$ label the ``internal" dimensions, we have 
$\r=\det E_{M'}^{~A'}$. This justifies the name ``dilaton" for $\r$.

By coordinate reparametrizations the $2d$ metric can be brought
into the conformal gauge at least locally:
\be
e_\m^{~\a}=\l\d_\m^{~\a}\equiv \exp(\s)\d_\m^{~\a}
\ee
In terms of the light-cone (isothermal) coordinates\footnote{We similarly
define $V^\pm:=V^0\pm V^1$ and $V_\pm := \ft12(V_0 \pm V_1)$ for any
vector $V^\m$.}
\ben
x^\pm := x^0\pm x^1,\qquad 
\p_\pm := \ft12(\p_0\pm\p_1)
\een
the metric takes the form
\be\la{confgauge}
{\rm d}s^2 = \exp(2\s)\; {\rm d}x^+{\rm d}x^-
\ee
It still admits conformal reparametrizations
$x^\pm\mapsto\tilde{x}^\pm(x^\pm)$, which preserve the diagonal form of the 
metric. Note that the Liouville degree of freedom $\s$ does not
transform as a genuine scalar. Rather, it is the following expression
\be\la{sh}
\sh\equiv \ln\l-\frac12\ln(\p_+\r\p_-\r),
\ee
which behaves as a scalar under conformal reparametrization
and which appears naturally in the equations of motion \ci{Nico91}.

\medskip \noindent
Let us now state the equations of motion for all fields:
\begin{itemize}
\item
The dilaton field $\r$ obeys a free field equation independently
of the coset $G/H$:
\be\la{eqmdilaton}
\Box \r = 0,
\ee
whose general solution is given by
\ben
\r(x)\equiv\ft12\big(\r^+(x^+)+\r^-(x^-)\big)
\een
The dual field (``axion") is defined by
\ben
\tilde{\r}(x)\equiv\ft12\big(\r^+(x^+)-\r^-(x^-)\big)
\een
\item  The conformal factor satisfies two (compatible) first order equations:
\be\la{eqmconf}
\p_\pm\r\p_\pm\sh = \ft12\r~\tr P_\pm P_\pm,
\ee
with $\sh$ defined above.  The equations \Ref{eqmconf} determine the
conformal factor up to a constant, since they are of first
degree. Rather than equations of motion of the usual type, they should
be regarded as constraints; actually they descend from variation of
the two unimodular degrees of freedom of the $2d$ metric $h_{\m\nu}$,
that appear as Lagrangian multipliers in \Ref{Lag42}. The second order
equation of motion for the conformal factor results from variation of
the Lagrangian w.r.t.~$\r$:
\be\la{eqmconf2}
\p_+\p_- \sh \equiv \p_+\p_- \sh = -\ft12\tr(P_+P_-)
\ee
\item  The matter fields $\cV$ obey
\be\la{eqmmatter}
D^\m(\r P_\m) = 2\big\{ D_+(\r P_-)+D_-(\r P_+) \big\} = 0
\ee
where the covariant derivative $D_\m\!=\!\p_\m+{\rm ad}_{Q_\m}$ was
introduced in the previous section. Similar equations appear in the
flat space $\s$-models (i.e.~\Ref{Lag42} without coupling to gravity)
except for the appearance of $\r$ here. The equations for the
conformal factor \Ref{eqmconf} further show that $\r$ may not be
chosen constant without trivializing the matter part of the
solution. This difference in \Ref{eqmmatter} accounts for the essentially
new features of these models in comparison with the flat space
models.
\end{itemize}
\noindent
Notice that all the equations of motion are consistent as
\Ref{eqmconf2} is a consequence of \Ref{eqmdilaton}, \Ref{eqmconf} and
\Ref{eqmmatter}, taking into account the relations \Ref{corel}.

There is an equivalent form of \Ref{eqmmatter}, frequently used in the
literature, in terms of the combination 
\be\la{g}
g \equiv \cV\n(\cV^{-1}) \in G,
\ee
(so e.g.~$g=\cV\cV^t$ for $SL(n,\R)$) which we state for later reference:
\be\la{eqmmatterg}
\p^\m(\r g^{-1}\p_\m g) ~=~ 2 \big\{ \p_+(\r g^{-1}\p_- g) 
+ \p_-(\r g^{-1}\p_+ g)\big\}  ~=~ 0
\ee
In the models of dimensionally reduced gravity the variables $g$
essentially build the compactified part of the former
higher-dimensional metric. Their main technical advantage, which we
will eventually exploit, is the fact, that in contrast to
\Ref{eqmmatter} the equations of motion \Ref{eqmmatterg} can be
formulated without $Q_\m$ (this is no longer true in the presence
of fermionic couplings). Indeed,
\be\la{geqm}
g^{-1}\p_\m g ~=~
 \n(\cV)\Big(Q_\m + P_\m -\n(Q_\m + P_\m) \Big)\n(\cV^{-1}) 
~=~ 2\n(\cV)P_\m \n(\cV^{-1})\nn
\ee

As an illustration let us once more consider the example
$G=SL(2,\R)$. This model is commonly formulated in terms of the
Ernst-potential $\E$, defined by
\be
\E = \D+iB
\ee
in the parametrization of \Ref{Vpar}. The equations of motion for the
matter part \Ref{eqmmatter} then take the form
\be
\D\p_\m(\r\p^\m\E) = \r\p_\m\E\p^\m\E,
\ee
which is the celebrated Ernst equation \ci{Erns68}. A further reduction
corresponding to $B\!=\!0$ leads to the collinearly polarized
so-called Einstein-Rosen gravitational waves. In this case the Ernst
equation reduces to the classical linear Euler-Darboux equation. These
solutions, discovered in \ci{EinRos37} have already served as testing
ground for technical and conceptual issues of quantum gravity
\ci{Kuch71,AshPie96a}.

\section{CLASSICAL INTEGRABILITY}
\setcounter{equation}{0}

\subsection{The Linear System} 
The main part of the equations of motion stated in the previous
section are the equations \Ref{eqmmatter} for the matter fields
$\cV(x)$. Once these equations are solved, the equations \Ref{eqmconf}
for the conformal factor can be integrated with \Ref{eqmmatter}
ensuring their integrability.

The matter field equations may be obtained from a linear system; this
means that they are expressed as compatibility equations of a linear
system of differential equations. These techniques have been common in
the theory of flat space integrable systems \ci{Fadd84,FadTak87}, where
the existence of a family of linear systems parametrized by the
spectral parameter in particular gives rise to the construction of an
infinite set of conserved charges, reflecting the integrability of
the model. For the axisymmetric stationary solutions of Einstein's
equations the linear system was constructed by Belinskii and Zakharov
\ci{BelZak78} and by Maison \ci{Mais78}. The generalization to
arbitrary non-linear $\s$-models is discussed in \ci{BreMai87,Nico91}. 
Alternatively, the integrability of these models can be derived
on the basis of  the (anti)self-dual Yang-Mills equations 
in four dimensions. This approach, which is quite different from 
the one taken here in that it emphasizes the twistor geometrical aspects 
has been extensively studied in \ci{selfdual1,selfdual2,selfdual3}. 

The linear system for a function $\vh(x,\g)$, where $\g$ denotes the
spectral parameter, is given by:
\ba
\vh^{-1}\p_\m\vh = Q_\m + \frac{1+\g^2}{1-\g^2}P_\m + 
\frac{2\g}{1-\g^2}\e_{\m\nu} P^{\nu} \la{ls}
\ea
or, in light-cone coordinates,
\ba
\vh^{-1}\p_\pm\vh =  Q_\pm + \frac{1\mp \g}{1\pm \g}P_\pm  \la{lslc}
\ea
This is formally almost the same linear system as for the flat space
$\s$-model. The essential difference lies in the fact, that in
order to obtain $D^\m(\r P_\m)\!=\!0$ as compatibility equations
rather than $D^\m P_\m\!=\!0$, the spectral parameter must depend on
the space-time coordinates according to the differential equation
\ba
&& \g^{-1}\p_\m\g=\frac{1+\g^2}{1-\g^2}\r^{-1}\p_\m\r + 
\frac{2\g}{1-\g^2}\e_{\m\nu}\r^{-1}\p^\nu\r \la{despectral}\\
&\Longleftrightarrow&
\g^{-1}\p_\pm\g=\frac{1\mp \g}{1\pm \g}\r^{-1}\p_\pm\r\nn
\ea
and is therefore no longer a constant as in the flat space integrable
systems.  This equation can be explicitly solved with two solutions
$\g, \g^*$ (due to the invariance of \Ref{despectral} with respect to
the involution $\g\!\mapsto\! 1/\g$):
\be\la{gamma}
\g(x,w) = \frac{\sqrt{w+\r^+(x^+)}-\sqrt{w-\r^-(x^-)}}
{\sqrt{w+\r^+(x^+)}+\sqrt{w-\r^-(x^-)}} = \frac1{\g^*(x,w)}
\ee
The free integration constant $w$ may be regarded as the hidden
``constant spectral parameter'' whereas $\g$ will be referred to as
the ``variable spectral parameter''. The reader should check that
compatibility of the linear system indeed requires \Ref{eqmmatter}.
\bigskip

The linear system \Ref{ls} determines the function $\vh(x,\g)$ only up to
constant left multiplication by an arbitrary matrix depending on $w$:
\be\la{mult}
\vh(x,\g)\mapsto S(w)\vh(x,\g)
\ee
We will restrain this freedom by further assumptions below.

To reconstruct the physical fields from $\vh$, we adopt a generalized
``triangular gauge'' for $\vh(x,\g)$ \ci{BJ2,BreMai87}, demanding
regularity for $\g\!\rightarrow\!0$.\footnote{Demanding regularity just at
$\g=0$ is, however, not quite sufficient to fix $\vh$ uniquely, see below.}
This allows the expansion
\ba
\vh(x,\g) &=& \exp\left[\varphi^{(0)}(x)+
\g\varphi^{(1)}(x)+\cO(\g^2)\right]\non
&=& \cV(x) + \cO(\g) \nn
\ea
Substituting this into the linear system \Ref{ls} yields the original
(non-linear) equations of motion for
$\cV(x)\!=\!\exp[\varphi^{(0)}(x)]$, i.e.~at $\g\!=\!0$. The next
order in $\g$ leads to
\ben
\p_\m\varphi^{(1)}=\e_{\m\nu}\p^\nu\varphi^{(0)} + 
\mbox{nonlinear terms},
\een
such that the fields $\varphi^{(1)}(x)$, $\varphi^{(2)}(x)$, \dots~
form an infinite hierarchy of dual potentials, analogous to
the one originally introduced in \ci{KinChi78} for the investigation of 
the action of the Geroch group on stationary axisymmetric solutions.
However, in order to properly implement the infinite
dimensional affine symmetries of the model, another infinity of
potentials $\varphi^{(-n)}, n\ge1$ is required; these symmetries are
extensively discussed in \ci{BJ2,BreMai87,Nico91, JulNic96}.
\bigskip

The involution $\n$ defining the symmetric space $G/H$ and acting on 
$\cV$ may be extended to an involution $\n^\infty$ acting on $\vh(\g)$ by
\ben
\n^\infty\Big(\vh(\g)\Big)\equiv\n\left(\vh\Big(\frac1{\g}\Big)\right)
\een
and leaving the linear system \Ref{ls} invariant. E.g.~for
$G=SL(n,\R)$, this means:
\ben
\n^\infty\Big(\vh(\g)\Big)=\left(\vh\Big(\frac1{\g}\Big)^t\right)^{-1}
\een
The existence of this generalized involution motivates the
following definition \ci{BreMai87}: 
\be\la{mon}
\cM \equiv\vh\n^\infty\!\left(\vh^{-1}\right) 
\quad
\left(~=\vh(\g)\vh\Big(\frac1{\g}\Big)^t~\right)
\ee
is called the {\em monodromy matrix} associated with $\vh(\g)$. Due to the
invariance of \Ref{ls} under $\n^\infty$, this matrix depends on the
constant spectral parameter only:
\ben
\p_\m\cM=0 \quad\Longrightarrow\quad \cM=\cM(w)
\een
Obviously, $\cM$ is not invariant w.r.t.~\Ref{mult}, but transforms as
\ben
\cM(w)\mapsto S(w)\cM(w)\n\left(S^{-1}(w)\right),
\een
s.t.~the form of $\cM$ may be restricted in various ways. Let us
comment on the two preferred choices (``pictures") eliminating the
freedom left by \Ref{mult}.
\begin{itemize}
\item{The freedom \Ref{mult} may be invoked to demand holomorphy of
$\vh(\g)$ inside a domain containing the unit disc
$|\g|\!\le\!1$. Roughly speaking, the invariance
$w(\g)\!=\!w(\g^{-1})$ allows to reflect all singularities at the unit
circle by multiplication with a suitable $S(w)$. This picture has been
introduced in \ci{BreMai87}. As a consequence, the matrix $\cM(w(\g))$
is non-singular as a function of $\g$ in an annular region containing
the unit circle $|\g|\!=\!1$ and contains the complete information
about $\vh$.  The linear system matrix $\vh(\g)$ may then be recovered
by solving a Riemann-Hilbert factorization problem on this
annulus. The absence of singularities in the disk in particular
permits us to recover the original field via $\cV (x) = \vh
(x,\g)|_{\g =0}$.}

\item{Since the matrix $\cM(w)$ obeys:
\be
\cM(w)=\n^\infty\!\left(\cM^{-1}(w)\right) 
\quad\Longrightarrow\quad \cM(w)=\n\left(\cM^{-1}(w)\right)
\ee
we can represent it as
\ben
\cM(w)=S(w)\;\n\!\left(S^{-1}(w)\right)
\een
Hence, by exploiting \Ref{mult} we can require
\be\la{BZ}
\cM(w)\equiv I,
\ee
for {\em all} solutions. This picture was introduced in \ci{BelZak78} and
will be used in the sequel. Note however, that it still
allows the freedom of left multiplication \Ref{mult} by $H$-valued
matrices $S(w)$ (for which $\n(S)\!=\!S$). The main difference with
the previous picture is that, while still regular at $\g=0$, the
linear system matrix $\vh$ is now allowed to have poles inside
the unit disk in the $\g$-plane as well.}
\end{itemize}
\bigskip

For later use, let us state an alternative version of \Ref{ls}.
Define $\Psi(\g)\!\equiv\!\vh(\g)\n(\cV^{-1})$, then
\ba
\Psi^{-1}\p_\pm\Psi &=& \n(\cV)\left(Q_\pm + 
\frac{1\mp \g}{1\pm \g}P_{\pm} - \n(Q_\pm+P_\pm)\right)\n(\cV^{-1}) 
\non
&=& \frac2{1\pm\g}\n(\cV)P_\pm\n(\cV^{-1}) \nn
\ea
This form of the linear system is tailored for the variables $g$
introduced in \Ref{g}, in terms of which it becomes
(cf.~\Ref{geqm})
\be\la{ls2}
\Psi^{-1}\p_\pm\Psi = \frac{1}{1\pm\g}\; g^{-1}\p_\pm g
\ee
The main advantage of this version of the linear system has already
been stressed above: the $Q_\pm$ do not appear explicitly. However, we
trade this technical simplification for the drawback of hiding part of
the group theoretical structure. This may already be seen from the
transformation behavior of $\cV$ \Ref{gauge}, that clearly exposes the
combined left and right action of the groups $G^{\mbox{\scriptsize
rigid}} \times H^{\mbox{\scriptsize local}}$.  The variables $g$ on
the other hand and consequently the $\Psi$-function transform under
the adjoint action of $G$ and remain invariant under $H$. For a proper
treatment of the coset structure, it will hence be necessary to return
to \Ref{ls}. This is essential when the action of the
solution-generating Geroch group is implemented by an
infinite-dimensional extension $G^\infty/H^\infty$ of the coset $G/H$
where the ``maximal compact subgroup" $H^\infty$ is defined by means
of the involution $\n^\infty$ \ci{BJ2,BreMai87,Nico91}.  The
original form of the linear system is also required for the extension
of the linear system to supergravity, since the $Q_\pm$ are
indispensable for the coupling to fermions. We will sketch this
generalization in the appendix.

Finally, we would like to stress that the conformal gauge \Ref{confgauge}
has been adopted for convenience only. It is, in fact, possible to
generalize the linear system to arbitrary (non-singular) $2d$ metrics
by means of the Beltrami differentials parametrizing the inequivalent
conformal structures of the world-sheets \ci{Nico94}. These constitute
extra physical albeit global degrees of freedom of the theory.

\subsection{Spectral Parameter Current}

So far, from the function $\vh$ we have considered only the defining
expressions $\vh^{-1}\p_\pm\vh$ which appear in the linear system
\Ref{ls}. As it turns out, however, the crucial quantity for our
subsequent considerations is the spectral parameter current
\be\la{spcurrent}
B(x;\g)~ := ~ \big(\vh^{-1}\p_\g \vh\big)(x,\g) ~\equiv~ 
\sum_{j=1}^N\frac{B_j(x)}{\g-\g_j(x)} ~\in ~ {\mathfrak g}_\C
\ee
The form of the right hand side defines the so-called isomonodromic
ansatz \ci{KorNic95a}; the $\g_j(x)$ are given by \Ref{gamma}:
$\g_j(x)\equiv\g(x,w_j)$. For later use we record the residue at
infinity
\be
B_\i := \sum_{j=1}^N B_j = \lim_{\g\rightarrow\i} \g B(\g) \la{Binfinity}
\ee
which governs the behavior of $\vh$ near $\g\!=\!\i$.  Observe that local
analyticity of the spectral parameter current as a function of $\g$
already follows from the linear system. The extra assumption in
\Ref{spcurrent} is that $B(\g)$ should be single-valued in the whole
$\g$-plane and possess only simple poles. These restrictions may be
justified by the fact that almost all known and physically interesting
axisymmetric stationary or colliding plane wave solutions of
Einstein's equations are of this type.\footnote{Higher order poles in
\Ref{spcurrent} would not only imply essential singularities in
$\vh(\g)$ but also rather nasty singularities in the actual
solutions.}  We will sketch the generalization to arbitrary spectral
parameter currents at the end of this chapter. This extension of the
formalism is also inspired by the treatment of the isomonodromic
solutions; naively the first step may be understood as replacing the
sum in \Ref{spcurrent} by an integral. Nevertheless, the hope is, that
the isomonodromic solutions as defined above are (in some sense) dense
in the ``space of all solutions''.
\bigskip

In the following, the spectral parameter current will be considered as
the fundamental object of the theory. Thus, in particular we should be
able to reconstruct the original currents $\cV^{-1}\p_\pm\cV$ from
knowledge of $B(\g)$. Indeed, writing out the derivatives $\p_\pm$ and
using \Ref{despectral} we get
\ba
\vh^{-1}\p_\pm\vh =  \left.\vh^{-1}\p_\pm \vh \right|_\g
    + \r^{-1}\p_\pm\r\frac{\g(1\mp \g)}{1\pm \g} 
\vh^{-1}\p_\g\vh 
\ea
Comparing this with \Ref{lslc}, taking $\g\!=\!\mp1$ and assuming
regularity of $\left.\vh^{-1}\p_\pm \vh\right|_\g$ at $\g\!=\!\mp1$
according to \Ref{spcurrent}, we infer that
\be\la{rel}
P_\pm(x) ~=~ \mp\r^{-1}\p_\pm\r \left. 
\vh^{-1}\p_\g \vh \right|_{\g=\mp1} 
~=~ \r^{-1}\p_\pm\r \sum_{j=1}^{N}\frac{B_j(x)}{1\pm\g_j(x)}
\ee
In the presence of fermions, regularity of the spectral parameter
current at $\g\!=\!\mp1$ is no longer valid and the argument must be
slightly modified, cf.~appendix.  Fixing the local $H$-gauge freedom
\Ref{gauge}, we can determine $Q_\pm$ as functions of $P_\pm$. In this
fashion, the complete current $\cV^{-1}\p_\pm\cV$ may be recovered
from \Ref{PQ}.  The (nontrivial) coset constraints ensuring
$P_\pm\!\in\!{\mathfrak k}$ will be discussed in section
\ref{Sconstraints} below.

Thus, we can now proceed to formulate the theory entirely in terms of
the new quantities $B(\g)$.
The former matter field equations of motion \Ref{eqmmatter} have been
completely absorbed into the linear system, i.e.~they are implicitly
part of the definition of the spectral parameter current $B(\g)$. This
already suggests that the coordinate dependence of this
current should be nothing but a consequence of its definition and its
$\g$-dependence. We shall explicitly work this out in the next section
where we will in particular study the $x^\pm$-dynamics of the residues 
$B_j$.

\subsection{Deformation Equations}

Let us investigate the consequences of the isomonodromic ansatz
\Ref{spcurrent}. For the derivation we make use of the alternative
form of the linear system given in \Ref{ls2} for which an analogous
spectral current may be defined:
\be\la{A}
A(x;\g) \equiv \Psi^{-1}\p_\g\Psi(x,\g) 
\equiv \sum_j\frac{A_j(x)}{\g-\g_j(x)},
\ee
The residues $A_j$ are obviously related to the $B_j$ by 
\be
A_j = \n(\cV)B_j\n(\cV^{-1}) \la{AB}
\ee
In analogy with \Ref{Binfinity} we define
\be
A_\i = \n(\cV)B_\i \n(\cV^{-1}) = \sum_{j=1}^N A_j \la{AB1}
\ee
The current \Ref{A} contains the original currents in a
way similar to \Ref{rel}:
\be\la{rela}
g^{-1}\p_\pm g ~=~ 
\mp2\r^{-1}\p_\pm\r \left. \Psi^{-1}\p_\g\Psi \right|_{\g=\mp1} 
~=~ 2\r^{-1}\p_\pm\r \sum_{j=1}^{N}\frac{A_j(x)}{1\pm\g_j(x)}\nn
\ee

Next we demonstrate in detail how the definition of the spectral
parameter current together with the isomonodromic ansatz uniquely
determines the $x^\pm$-dynamics of the residues $A_j$ and $B_j$.
The definition \Ref{A} requires validity of the compatibility
equations:
\be\la{compa}
\Psi^{-1}\left[\p_\pm,\frac{\p}{\p\g}\right]\Psi ~=~
\p_\pm\left(\Psi^{-1}\frac{\p\Psi}{\p\g}\right)
- \frac{\p}{\p\g}\left(\Psi^{-1}\p_\pm\Psi\right)
+ \left[\Psi^{-1}\p_\pm\Psi~,~\Psi^{-1}
\frac{\p\Psi}{\p\g}\right]
\ee
Due to the explicit coordinate dependence of the variable spectral
parameter $\g$, the l.h.s.~does not vanish, but is
\ben
\left[\p_\pm,\frac{\p}{\p\g}\right] = \p_\pm\left(\frac{\p w}{\p\g}\right) 
\frac{\p}{\p w}
= \r^{-1}\p_\pm\r\left(1-\frac2{(1\pm\g)^2}\right)\frac{\p}{\p\g}
\een
Together with the linear system \Ref{ls2} and \Ref{rela} this permits
us to express the compatibility relations \Ref{compa} entirely in terms
of the spectral parameter current $A(\g)$:
\be\la{compA}
\underbrace{\p_\pm A(\g)}_{\rm l.h.s.} = \r^{-1}\p_\pm\r
\left\{\underbrace{\frac{2A(\mp1)}{(1\pm\g)^2}}_a
+\underbrace{\left(1-\frac2{(1\pm\g)^2}\right)A(\g)}_b
\pm\underbrace{\frac2{1\pm\g}\left[A(\mp1),A(\g)\right]}_c\right\}
\ee
We evaluate the l.h.s.~and the three terms of the r.h.s.~separately:
\ba
\mbox{l.h.s.} &=& \p_\pm\left(\sum_j\frac{A_j}{\g-\g_j}\right) \non
&=& \sum_j\frac1{\g-\g_j}\p_\pm A_j 
    + \r^{-1}\p_\pm\r \sum_j\frac{A_j}{(\g-\g_j)^2}
      \left\{\frac{\g_j(1\mp\g_j)}
{1\pm\g_j}-\frac{\g(1\mp\g)}{1\pm\g}\right\}\non
&=& \sum_j\frac{\p_\pm A_j}{\g-\g_j} ~+~ \r^{-1}\p_\pm\r 
\sum_j\frac{A_j}{\g-\g_j}\left\{1-\frac2{(1\pm\g)(1\pm\g_j)}\right\} 
\la{z1}\\
&&\non
&&\non
a &=& 
\mp\frac2{(1\pm\g)^2}\left( 
\sum_j\frac{A_j}{1\pm\g_j} \right) \non
&=& \frac2{1\pm\g}\r^{-1}\p_\pm\r \sum_j
\frac{A_j}{\g-\g_j}\left\{\frac1{1\pm\g}-\frac1{1\pm\g_j} \right\}\\
&&\non
&&\non
b &=& 
\left(1-\frac2{(1\pm\g)^2}\right)
\sum_j\frac{A_j}{\g-\g_j}\\
&&\non
&&\non
c&=& -\frac2{1\pm\g}\left[\sum_j\frac{A_j}{1\pm\g_j}~,~
\sum_k\frac{A_k}{\g-\g_k}\right] \non
&=& -\frac2{1\pm\g} \sum_{j,k}
\frac{[A_j,A_k]}{(1\pm\g_j)(1\pm\g_k)} \non
&=& \frac1{1\pm\g}\sum_{j,k}[A_j,A_k] 
\frac{(1\pm\g)(\g_j-\g_k)}{(1\pm\g_j)(1\pm\g_k)(\g-\g_j)(\g-\g_k)} \non
&=& 2\sum_{j,k} \frac{[A_j,A_k]}{(1\pm\g_j)(1\pm\g_k)}
\frac1{\g-\g_j}\la{z4}
\ea
\bigskip

\noindent
Combining \Ref{z1}--\Ref{z4} now yields the deformation equations 
for the residues $A_j$ \ci{KorNic95a}
\be\la{defA}
\p_\pm A_j = 2 \r^{-1}\p_\pm\r \sum_{k} 
\frac{[A_j,A_k]}{(1\pm\g_j)(1\pm\g_k)}
\ee
The corresponding deformation equations for the residues $B_j$ are
easily derived from this by use of \Ref{AB}
\ba
D_\pm B_j &=& 2 \r^{-1}\p_\pm\r \sum_{k} 
\frac{[B_j,B_k]}{(1\pm\g_j)(1\pm\g_k)} 
-[B_j,P_\pm]\la{defB}\\
&=&\r^{-1}\p_\pm\r \sum_{k}
\frac{1\mp\g_j}{(1\pm\g_j)(1\pm\g_k)} [B_j,B_k] \nn
\ea
Again, we note the explicit appearance of the $Q_\pm$, which are part
of the covariant derivative $D_\pm\!=\!\p_\pm\!+\!\mbox{ad}_{Q_\pm}$,
whereas $P_\pm$ has been expressed in terms of $B(\g)$ according to
\Ref{rel}.  We may also translate the deformation equations
\Ref{compA} into the spectral parameter current $B(\g)$:
\ba
D_\pm B(\g) &=& \rho^{-1}\p_\pm\rho 
\left\{\frac{2B(\mp1)}{(1\pm\g)^2} 
+ \left(1-\frac2{(1\pm\g)^2}\right)B(\g) 
\pm\frac2{1\pm\g}\left[B(\mp1),B(\g)\right]\right\}\non
&&{}-\left[B(\g),P_\pm\right]\la{compB}
\ea

Together with the equations for the $\g_j$ following from
\Ref{despectral}, the dynamics of the spectral parameter currents
$A(\g)$ and $B(\g)$ in the $x^{\pm}$ directions is completely given
from \Ref{defB}. Remarkably, these deformation equations are
automatically compatible; we leave this computation as a little
exercise to the reader.

In summary, we have reduced the Einstein equations for the stationary
axisymmetric or the colliding plane wave truncation and their
generalizations to arbitrary $\s$-models to a system of compatible
ordinary first order matrix differential equations. Any solution of
\Ref{defA} induces a solution of the matter field equations of motion
\Ref{eqmmatter} from which the conformal factor may be determined by
integration of \Ref{eqmconf}. The simplest class of solutions is
obtained if the $A_j$ are restricted to the Cartan-subalgebra of
$\mathfrak{g}$; then, they are in fact constant owing to
\Ref{defA}. This special case includes for instance the
multi-Schwarzschild solutions and illustrates the remarkable fact that
the coordinate dependence of these solutions arises solely through the
coordinate dependence of the spectral parameters $\g_j (x)$.

The form of \Ref{defA} gives rise to several constants of motion.
{}From \Ref{defA} one immediately verifies the conservation laws
\ci{KorNic96}
\be
\p_\pm A_\i = \p_\pm (\tr A_j^n) = 0 \qquad (n=1,2,\dots) \la{conservation}
\ee
which have obvious analogs in terms of the $B_j$ variables.  We call
$A_\i$ the ``Ehlers charge" because it generates (the analog of)
Ehlers transformations in the canonical approach \ci{KorNic96}. The
constancy of the traces tells us that the eigenvalues of the matrices
$A_j$ (or $B_j$) are likewise constant (note that $\tr A_j$ vanishes
for semisimple $G$). Later on we shall also describe a more general
construction of conserved charges non-local in the spectral parameter
plane.

The above reduction of the original equations shows a remarkable
general feature: the number of dimensions has been effectively reduced
from two to one. Recall that the initial values of the physical fields 
are usually given on a spacelike hypersurface, whereas their
evolution in the time direction is described by the equations of
motion. Here, on the contrary we have evolution equations for the time
direction as well as for the space direction and the two flows
commute. The knowledge of the initial values of the new fields $B_j$
and $\g_j$ at one space-time {\em point} is sufficient to reconstruct
the whole solution by means of \Ref{defB} and \Ref{despectral}.

This at first sight puzzling feature may be understood as follows: the
space dimension which previously provided the initial data has been
traded for an additional dimension parametrized by the spectral
parameter.  In fact, given the spectral parameter current $B(\g)$ at
fixed $\g\!=\!\pm1$ on a spacelike hypersurface (which according to
\Ref{rel} are nothing but the original currents) allows us to evolve
it into time direction by means of the equations of motion and into
the $\g$-direction via the compatibility equations \Ref{compB}. Vice
versa, given $B(\g)$ at fixed space-time point but for all $\g$ one
can deduce its space evolution from the compatibility equations. The
isomonodromic ansatz is finally employed to parametrize the behavior
of the spectral parameter current in the $\g$-plane by a discrete
(even finite) set of variables, such that the original field theory
reduces to an ``$N$-particle'' problem (note however, that these
``particles'' are localized only in the spectral parameter plane, but
{\em not} in the actual space-time and should accordingly be referred
to as ``$N$-waves''). In this way we have arrived at an effectively
one-dimensional description of the $2d$ theory without sacrificing the
nontriviality of the solutions. In particular, the quantization of
this system will resemble that of an ordinary quantum mechanical
system. From the point of view of quantum field theory, one may think
of it as a kind of collective quantization.

\subsection{Reality and Coset Constraints}
\label{Sconstraints}
We have introduced the new variables $A(\g)$ or equivalently $B(\g)$
as fundamental objects. However, in the way they have been defined,
these variables contain too many degrees of freedom.  In order to
restore the original fields $g$ and $\cV$, respectively, from the
spectral parameter currents, additional constraints must be imposed in
order to ensure the reality and coset properties of the original
variables, such that e.g.~$\cV\!\in\!G$ rather than
$\cV\!\in\!G_\C$. For the sake of clarity of presentation we will only
consider the spectral parameter current $B(\g)$, in terms of which the
reality and coset conditions can be stated most clearly. These
conditions are direct consequences of \Ref{rel} and the corresponding
constraints on $P_\pm$.

The first important observation is that, for complex $\g\!\in\!\C$,
$B(\g)$ lives in the {\em complexified} Lie algebra $\mathfrak{g}_\C$. 
To ensure reality of the physically relevant quantities (in
particular of $B(\pm1)$) we impose the constraint
\be\la{reality}
\overline{ B(\g)} = B(\gb)
\ee
Here the bar denotes complex conjugation on the complexified
Lie algebra $\mathfrak{g}_\C$ defined by 
\be
\overline{\sum_a y_a Z^a} := \sum_a \overline{y_a} Z^a  
   \;\;\;\;  ( y_a\in\C )  \la{reality1}
\ee
where the generators $Z^a$ span the chosen real form $\mathfrak{g}$. 
Let us remark already here that the appearance of the complexified
Lie algebra is very important for the quantum theory 
since the structure of unitary irreducible representations 
of $\mathfrak{g}_\C$ is quite different from that of $\mathfrak{g}$.

{}From \Ref{rel} we see that for $P_\pm\in{\mathfrak{k}}$, we must demand
in addition that
\be
B(\pm1)\in\mathfrak{k}\subset\mathfrak{g},
\ee
Thus, $B(\pm1)$ is constrained to take values in the
non-compact part of the real Lie algebra $\mathfrak{g}$. 
A priori this is not a consequence of the isomonodromic ansatz
since the generic solution of the deformation equations will
not satisfy this constraint.

To further analyze the coset constraint recall that we work 
in the BZ-picture where $\cM =I$ \Ref{BZ}, whence 
\ben
0 = \frac{\p \cM}{\p w} \frac{\p w}{\p \g} 
     =\vh(\g)\left\{B(\g)+\frac1{\g^2}\;
\n\left(B\left(\frac1{\g}\right)\right)\right\}~\n\left(\vh^{-1}
\left(\frac1{\g}\right)\right), 
\een
i.e. 
\be\la{coset}
B(\g)+\frac1{\g^2}\;\n\left(B\left(\frac1{\g}\right)\right) = 0.
\ee
This constraint immediately leads to
\be\la{coset1}
B(\pm1)+\n(B(\pm1))=0 \quad\Longrightarrow\quad  B(\pm1)\in\mathfrak{k}
\ee
Moreover, multiplying \Ref{coset} by $\g$ and taking $\g\ra\i$ 
we see from \Ref{Binfinity} that \Ref{coset} implies
\be
B_\i = - \lim_{\g\ra 0} \g \n\big( B(\g)\big) \la{Binfnull}
\ee
which vanishes if $B(\g)$ is regular at $\g =0$ (i.e., in the
triangular gauge). Conversely, for $B_\i \neq 0$, we see that $B(\g)$ 
must have a first order pole at $\g =0$ with residue $-\n(B_\i)$.

The reality and coset constraints \Ref{reality} and \Ref{coset}
effectively reduce the number of degrees of freedom of the spectral
parameter current. In terms of the isomonodromic ansatz
\Ref{spcurrent} we can satisfy them for instance by taking $N\!=\!2n$
and real poles $\g_j$ with:
\be
\g_j = \frac1{\g_{j+n}} \in \R \quad , \quad 
B_j = \overline{B_j} = \n(B_{j+n}) = \overline{\n(B_{j+n})}\qquad
\big( j=1,\dots,n \big)
\la{polreal}
\ee
and
\be
\sum_{j=1}^{n}\Big(B_j+\n(B_j)\Big) = 0 \quad\Longleftrightarrow\quad
\sum_{j=1}^{n} B_j \in {\mathfrak k}  \la{cosreal}
\ee
In this case all residues belong to the real form of the Lie algebra,
i.e.~$B_j\in{\mathfrak g}$.  Alternatively, and more generally, we can
take the number of (complex) poles to be a multiple of four,
i.e.~$N\!=\!4n$ with
\be\la{polrea}
\g_j=\gb_{j+n}=\frac1{\g_{j+2n}}=\frac1{\gb_{j+3n}}\in\C
\qquad \big( j=1,\dots,n \big)
\ee
and
\be\la{polcos}
B_j = \overline{B_{j+n}} = \n(B_{j+2n}) = \n(\overline{B_{j+3n}}) 
\ee
Now \Ref{Binfnull} is ensured by
\be
\sum_{j=1}^{2n} \big( B_j + \n (B_j) \big) = 0 \quad\Longleftrightarrow\quad
  \sum_{j=1}^{2n} B_j \in {\mathfrak k}
\ee
Consequently, we now have $B_j\in {\mathfrak g}_\C$. Of
course, we can also consider both possibilities together, when only
some of the poles and residues are subject to \Ref{polreal}, and the
remaining ones satisfy \Ref{polrea} together with \Ref{polcos}.  

We leave it as an exercise to the reader to check that the deformation
equations \Ref{defB} are indeed compatible with all these constraints
(taking into account that $\n$ and complex conjugation are
automorphisms).  Furthermore, the reader may reformulate the complete
set of constraints in terms of the alternative spectral parameter
current $A(\g)$ to convince himself that the coset constraints take a
far less convenient form there due to the explicit appearance of the
original fields $\cV$ or $g$ in the constraints.

\subsection{$\t$-Function and Conformal Factor}
Having resolved the matter part of the model, we are still left
with the first order equations of motion \Ref{eqmconf} for the 
conformal factor. These formulas relating
the gravitational to the matter sector have a well known analog
in the theory of integrable models. 
Though there is no conformal factor in the flat space
models, an additional function can be introduced by these
equations, which contains the complete information 
about the system in the sense that it serves as generating 
function for the Hamiltonians of the corresponding deformation 
dynamics. This is the so-called $\t$-function.

Let us explain this in more detail. Define the one-form
\be\la{omega}
\o_0\equiv\frac12\sum_{j\not=k}\tr(A_jA_k)~{\rm d}\ln(\g_j-\g_k),
\ee
which was originally introduced in \ci{JiMiMoSa80,JiMiUe81}. It
is usually understood as a function of the parameters $\g_j$ and turns
out to be closed
\ben
{\rm d}\o_0 = 0,
\een
if the residues $A_j$ (considered as functions of the $\g_k$) are 
subject to the so-called Schlesinger equations
\be\la{Schles}
\frac{\p A_k}{\p\g_j} = \frac{[A_j,A_k]}{\g_j-\g_k}\quad ,\quad
\frac{\p A_j}{\p\g_j} = -\sum_{k\not=j}\frac{[A_j,A_k]}{\g_j-\g_k},
\ee
which describe a $\g_j$-dependence induced by certain isomonodromy
conditions. This gives rise to the definition of the so-called
$\t$-function:
\be\la{tau}
{\rm d}\t=\o_0
\ee
This function generates the Hamiltonians governing the
$\g_j$-dynamics of \Ref{Schles} \ci{JiMiMoSa80}.
\bigskip

We will now show how to apply this concept to our model, 
treating all quantities as functions of the coordinates $x^{\pm}$. In
particular, we will assume the dependence $A_j=A_j(\{\g_k(x^{\pm})\})$,
i.e.~the residues depend on the coordinates $x^{\pm}$ only via the
$\g_k$. It will turn out, that the corresponding $\t$-function equals
the conformal factor up to some explicit factor. It the next section
we will further see that the Hamiltonians for the $x^{\pm}$-dynamics
arise in the same way.

Let us again define the one-form $\o_0$ by \Ref{omega}, but now
with respect to the space-time coordinates $x^\pm$,
taking into account the coordinate-dependence of the $\g_j$:
\be
\o_0 = \frac12\sum_{j\not=k} \frac{\tr(A_jA_k)}{\g_j-\g_k}
\Big[\p_+(\g_j-\g_k){\rm d}x^+ + \p_-(\g_j-\g_k){\rm d}x^-\Big]
\ee
We leave it as an exercise to show, that again $\o_0$ is closed if the
residues $A_j$ satisfy the deformation equations \Ref{defA}. Thus, the
$\t$-function may be defined as in \Ref{tau}, the differential again
taken w.r.t.~$x^\pm$. Notice that it does not matter for
this definition which of the spectral parameter currents $A$ and $B$ 
we employ, as they only differ by matrix conjugation which drops
out in the trace.

Next we observe that the equations of motion for the conformal factor
take the following form upon substitution of \Ref{rel} into \Ref{eqmconf}
\be
\p_\pm\r\p_\pm\sh = \ft12\r~\tr P_\pm P_\pm = 
\ft12\r^{-1}(\p_\pm\r)^2\sum_{j,k}
\frac{\tr(A_jA_k)}{(1\pm\g_j)(1\pm\g_k)} \la{eqmconf1}
\ee
which we rewrite as 
\ba
\p_\pm\sh ~=~ \r^{-1}\p_\pm\r && \hspace{-1em}
\left\{~\frac12\sum_{j\not=k}\tr(A_jA_k)
\left[\frac2{(1\pm\g_j)(1\pm\g_k)}-1\right] \right. \non
&&{}\left.+\sum_j\frac{\tr A_j^2}{(1\pm\g_j)^2}
+\frac12\sum_{j\not=k}\tr(A_jA_k) ~\right\}\nn
\ea
The first term on the r.h.s.~is just $\p_\pm\ln\t$. 
For the second line we use
\ben
\sum_{k\not=j}\tr(A_jA_k) = 
\tr A_\infty^2- \tr A_j^2,
\een
which is constant by \Ref{conservation}. Thus, the corresponding
terms in the above equation can be integrated explicitly; we have
\ba
\p_\pm\sh &=& \p_\pm\t + \r^{-1}\p_\pm\r\left\{\sum_j\tr A_j^2
\left(\frac1{(1\pm\g_j)^2}-\frac12\right) +\frac12\tr A_\infty^2\right\}
\non
&=& \p_\pm\t + 
\sum_j\frac12\tr A_j^2~\p_\pm\left(\ln\frac{\p\g_j}{\p w_j}\right) 
+\frac12\tr A_\infty^2\p_\pm\left(\ln\r\right)\nn
\ea
which may be integrated up to give
\be\la{tauconf}
\hat{\l} = \exp\sh = 
\r^{\frac12 \tr A_\infty^2}\prod_{j=1}^N
\left(\frac{\p\g_j}{\p w_j}\right)^{\frac12\tr A_j^2} ~\cdot ~\t
\ee
Thus the conformal factor coincides with the $\t$-function up to an
explicit factor. Note that for solutions satisfying the coset
constraints the first factor is unity since $A_\i =0$; this factor
would, however, play a role for axisymmetric stationary solutions
which are not asymptotically flat, but have a non-vanishing Kasner
parameter. If moreover the whole set of coset constraints
\Ref{polreal} is fulfilled, the relation \Ref{tauconf} takes the
form:
\be\la{tauconfcos}
\hat{\l} =  
\prod_{j=1}^n
\left(\g_j^{-1}\frac{\p\g_j}{\p w_j}\right)^{\tr A_j^2} ~\cdot ~\t
\ee

In the next chapter we will show that the conformal factor in 
these models does have all the requisite properties of
a $\t$-function in that it serves as a generating function 
for the various Hamiltonians. The relation \Ref{tauconf} will 
reappear in the quantized theory and reveal an intriguing link between 
our model and conformal field theory. 

\subsection{Beyond the Isomonodromic Sector}
\la{beyiso}
The restrictions implied by the isomonodromic ansatz \Ref{spcurrent}
can actually be relaxed. In this section, we sketch how to
generalize the treatment to arbitrary non-singular classical
solutions.  One of the main advantages of the isomonodromic ansatz was
the possibility to parametrize the spectral parameter current $B(\g)$
by a complete set of quantities with rather simple evolution equations
\Ref{defB}, whereas the deformation equations \Ref{compB} in terms of
a general $B(\g)$ took a rather complicated form. The task at hand is
then to identify a proper analog of the variables $B_j$ in the general
case with deformation equations similar to \Ref{defB}.

\begin{figure}[htbp]
  \begin{center}
    \leavevmode
     \input{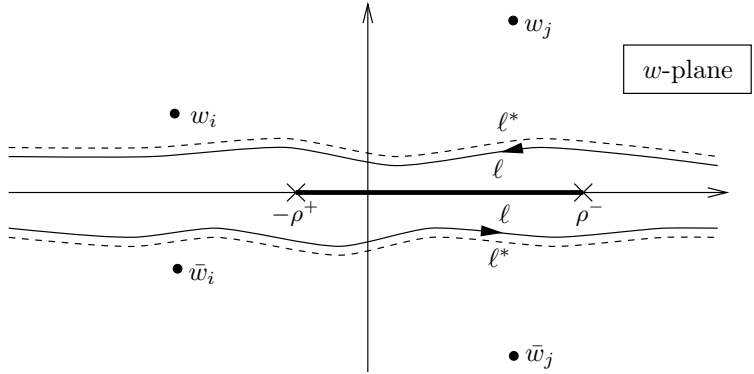}
  \end{center}
  \caption{Paths $\ell$ and $\ell^*$ in the $w$-plane}
  \label{l1}
\end{figure}

Let us assume that the function $B(x;\g(x,w))$ is holomorphic as a
function of $w$ in a small striplike region parallel to the real axis
of the complex $w$-plane on both sheets of the Riemann surface of the
function $\sqrt{(w\!+\!\r^+(x^+))(w\!-\!\r^-(x^-))}$, which has a
``movable" (i.e.~$x$-dependent) branch cut extending from $-\r^+(x^+)$
to $\r^-(x^-)$ along the real axis. We will denote this double strip
by $D$ and its (oriented) boundaries by $\ell \cup \ell^* \equiv \p
D$; by assumption there exists $\varepsilon >0$ such that the distance
from $\p D$ to the real axis is always greater than $\varepsilon$. 
The image of the double strip $D$ in the $\g$-plane
\be
\cD (x) := \big\{ \g\in\C \, | \, \g=\g(x;w), w\in D\big\}
\ee
is a ``movable" annular region containing the unit circle $|\g|=1$.
Since the exchange of the two sheets in the $w$-plane corresponds to
the involution $\g\rightarrow 1/\g$ in the $\g$-plane, the two
contours $\ell_\g$ and $\ell_\g^*$ bounding $\cD$ are related in the
same way. These contours are illustrated in figures \ref{l1} and
\ref{l2}.
\begin{figure}[htbp]
  \begin{center}
    \leavevmode
     \input{bey2.pstex_t}
  \end{center}
  \caption{Paths $\ell_\g$ and $\ell^*_\g$ in the $\g$-plane}
  \label{l2}
\end{figure}

The assumed regularity of $B(x;\g(x,w))$ in the whole double strip $D$
implies that the singularities never meet the branch cut, and hence
the corresponding solutions of the original field equations are
non-singular in space-time (as is, for instance the case for
Einstein-Rosen waves). In the $\g$-plane this means that that the
movable singularities always stay away from the annulus $\cD(x)$, and
hence from the circle $|\g|=1$.

In order to generalize the ansatz \Ref{spcurrent} to non-isomonodromic
solutions we observe that with the above assumptions we can
write the former as
\be\la{Cauchy}
B(x;\g) = \oint_{\p \cD} 
\frac{\cF(x,\tilde \g)}{\tilde \g-\g}\frac{{\rm d}\tilde \g}{2\pi i}
\ee
by Cauchy's formula where $\cF$ coincides with the
boundary values of $B$. We can now exploit the freedom of adding
any function holomorphic outside of $\cD$ to demand that
\be
\cF(x,\tilde \g) = 2\pi i\; \B (x;w(x,\tilde \g))
\frac{\p w(x;\tilde \g)}{\p \tilde \g}
\ee
where $\B$ is defined to have the $x$-dependence 
\be\la{bdep}
D_\pm\B(w)=
2\r^{-1}\p_\pm\r \oint_{\ell\cup \ell^*}
\f{[\B(w),\;\B(v)]}{(1\pm\g(w))(1\pm\g(v))} \; {\rm d} v 
- [\B(w),\;P_\pm],
\la{DEFB}\ee
such that in particular $\cF$ does explicitly not coincide with the
boundary values of $B$ any longer. For consistency (as we could of
course not demand any form of \Ref{bdep}), it must be ensured, that
\Ref{DEFB} via \Ref{Cauchy} still implies the deformation equations
\Ref{compB}, which indeed it does. Summarizing, we have
\be\la{Cau}
B(x;\g) =
    \oint_{\p \cD(x) } \frac{\B (x;w(x,\tilde \g))}{\tilde \g -\g} 
    \frac{\p w(x;\tilde \g)}{\p \tilde \g}\; {\rm d}\tilde \g
= \oint_{\p D}  \frac{\B (x;w)\; {\rm d}w}{\g (x;w) -\g} 
\ee
Comparing \Ref{DEFB} with the deformation equations of the residues
\Ref{defB}, we recognize $\B$ as the proper generalization of
the $B_j$ from the isomonodromic sector.
In the limiting case, where the singularities $\g_j$ just lie on the
contour $\p\cD$, we can embed the
isomonodromic sector into the general framework by setting
\be
\B(w)=-\sum_{j=1}^{N} B_j \delta (w-w_j)\;\;\;\;\;\;\;
w,\;w_j\in \ell
\la{Biso}\ee
where 
\be
\delta(w\!-\!w_j) := \frac{ds}{dw} \d(s-s_j)
\ee
with some affine parameter $s$ along the curve $\p D$ (note that
this definition is independent of the chosen parametrization).

It remains to generalize the reality and coset constraints. This is 
easily done: the reality constraint \Ref{reality} reads
\be
\overline{\B (x;w)}  = \B(x; \bar{w})
\ee
The coset constraint \Ref{coset}, on the other hand is ensured by
demanding 
\be
\B(x;w)=\n\left(\B(x;w^* )\right)
\qquad\mbox{for}\quad  w\in \p D
\la{cB}\ee
where $w^*$ is the point lying over $w$ in the second sheet.

\section{HAMILTONIAN FORMULATION}
\setcounter{equation}{0}
The form of the deformation equations \Ref{defB} suggests that we
should seek a Hamiltonian formulation of the theory which exploits
the decoupling exhibited by these equations and look for {\em two} 
Hamiltonians (or more precisely for two Hamiltonian constraints) describing 
the evolutions in $x^\pm$-directions, respectively. Evidently, this would
not be quite the usual Hamiltonian formalism where time is singled out
as the direction in which ``something happens" and where
the canonical variables depend on the coordinate $x^1$ of some
fixed space-like (one-dimensional) hypersurface.\footnote{We note
that reservations with regard to the special role of time in the usual 
canonical formalism were expressed already long ago by Dirac \ci{Dira49}.}
Rather, the canonical variables in our formulation depend on
the spectral parameter $\g$, such that the brackets are given as
``equal point brackets" at a fixed space-time reference point
instead of the usual equal time brackets containing spatial 
$\d$-functions. One obvious advantage of this formulation is
its $2d$ covariance which is manifest at every stage.

Although the relation of these ideas with the standard formalism
remains to be fully elucidated, we will first review the conventional
Hamiltonian treatment of the models in the following
section.\footnote{See also \ci{Hus96} for a recent discussion of
symmetry reduced Einstein gravity in the framework of Ashtekar's
variables and \ci{Mena96} for the quantization of Levi-Civit\`a
spacetimes.}  For the truncation corresponding to Einstein-Rosen waves
this program can be carried to completion \ci{Kuch71,AshPie96a}.
However, for the full theory with (non-polynomially) self-interacting
matter fields the conventional treatment gets bogged down in technical
difficulties which have not yet been completely overcome even for the
simpler flat space non-linear $\s$-model. To obviate this impasse new
methods are evidently needed, and our ``two-time formalism" addresses
precisely these difficulties, exploiting many results of the theory of
flat space integrable systems \ci{Fadd84,FadTak87}.  A further
motivation for presenting both formulations side by side is to
underline the similarities between the present model and string
theory, which can be viewed as the simplest example of a solvable
model of matter coupled quantum gravity in two dimensions. In
particular, the notorious ``time problem" of quantum gravity
\ci{Isham94} was already considered (and solved) in this context by
string theorists a long time ago. The choice of Weyl canonical
coordinates for axisymmetric stationary gravity, or its Lorentzian
analog, precisely corresponds to the light-cone gauge in string
theory.

In canonical quantum gravity it is customary to
make certain assumptions about the global structure of space-time 
(existence of a global time coordinate, asymptotic flatness, etc.) 
in order to disentangle gauge from dynamics and
to extract from the WDW equation a Schr\"odinger-type
equation for the matter degrees of freedom (see also \ci{Kief93} for a
discussion and further references on this topic). To be sure, such
assumptions are indispensable if one wants the quantum theory to
possess standard Fock space properties (distinguishability of positive and
negative frequencies, existence of a vacuum).
However, we would prefer to avoid them here because 
Fock space concepts are clearly not of much use in the presence
of non-polynomial interactions, and furthermore do not exploit
the underlying symmetry structures of the model.
For this reason, we will set up the theory 
locally and restrict attention to a particular coordinate patch, 
separating the gauge part of the evolution
(bubble time evolution) from the dynamics as a local observer would
do.\footnote{The intuitive idea behind this procedure is that 
an earth based quantum gravity practioner should not need to 
make any assumptions about the structure of space-time for an 
observer in Andromeda galaxy in order to be able to extract a Schr\"odinger 
equation from the WDW equation, if he assumes all his matter 
wave functions to vanish beyond (say) Pluto's orbit.} An extra bonus
is that this description can be naturally extended to more complicated
space-time topologies, i.e.~arbitrary (Lorentzian) Riemann surfaces.

\subsection{Conventional Approach}
We will not use the original Lagrangian \Ref{Lag42}, because for the canonical 
formulation it is somewhat more convenient to work with an equivalent
one presented in \ci{JulNic96} that treats the dilaton $\r$ 
and the axion $\rt$ as independent fields, and which reads 
\be
\cL = \et^{\m \nu} \o_\m \pa_\nu \r + \gt^{\m\nu} \o_\m \pa_\nu \rt
      + \ft12 \r \gt^{\m\nu} {\rm Tr} P_\m P_\nu \la{Lagr}
\ee
with $\et^{\m\nu}:= e e_\a^{\;\; \m} e_\b^{\;\;\nu} \e^{\a\b}$
and $\gt^{\m\nu}:= e e_\a^{\;\;\m} e_\b^{\;\;\nu} \eta^{\a\b}$.
Since $\et^{\m\nu}$ is a density and
the conformal factor drops out from the unimodular metric
$\gt_{\m\nu}$, \Ref{Lagr} is manifestly Weyl invariant. 
Let us first check that this Lagrangian gives rise to the desired 
equations of motion. Varying $\o_\m$ yields
\be
\et^{\m\nu}\p_\nu \r + \gt^{\m\nu} \p_\nu \rt =0  \la{eqdil}
\ee 
which implies (the generally covariant analog of) \Ref{eqmdilaton}.
The variation of $\rt$ tells us that $\o_\m$ is a curl, i.e.
there exists a scalar $\sh$ such that
\be
\o_\m=\et_\m^{\;\;\;\nu} \p_\nu \sh \la{conffact}
\ee 
Subsequent variation of $\r$ and use of \Ref{conffact} leads to
\be
\p_\m(\gt^{\m\nu} \p_\nu \sh)= - \ft12 \gt^{\m\nu} {\rm Tr} P_\m P_\nu
\ee
This is just our previous equation \Ref{eqmconf2}; therefore the field
$\sh$ is indeed the same as the one introduced in \Ref{sh}.
Finally, variation of $\gt_{\m\nu}$ 
together with \Ref{conffact} reproduces the two first 
order equations for the conformal factor \Ref{eqmconf}.

The formal disappearance of the conformal factor from the Lagrangian
\Ref{Lagr} can also be seen in the canonical framework. Let us first
compute the canonical momenta of the non-matter degrees of freedom
\ci{JulNic96}
\ba
\Pi:= \frac{\d \cL}{\d \pa_0 \r} = \o_1 \;\;\; , \;\;\;
\Pit:= \frac{\d \cL}{\d \pa_0 \rt} = \gt^{00} \o_0 
  + \gt^{01} \o_1 \;\;\; , \;\;\;
\O^\m:= \frac{\d \cL}{\d \pa_0 \o_\m} = 0  \la{canmom1}
\ea
Hence, by \Ref{conffact},
\be 
\Pi = -\p_0 \sh \;\;\; , \;\;\; \Pit = -\p_1 \sh  \la{Pisigma}
\ee
The four equations \Ref{canmom1} constitute a set of second class
constraints.  Following standard procedures \ci{Dira67,HenTei92} we
can thus eliminate the variables $\o_\m$ and $\O^\m$, keeping only
$\r$, $\rt$ and their canonical momenta as independent phase space
variables.  The relevant Dirac brackets are then given by
\ba
\{ \Pi (x), \r(y) \} &=& \{ \Pit (x) , \rt(y) \} = \dxy  \non
\{ \Pi (x), \rt(y) \}&=&\{ \Pi (x) , \rt(y) \} = 0  \la{Dirac1}
\ea
where from now on in this section $x\equiv x^1, y\equiv y^1,...$
denote {\em spatial} coordinates at fixed time $x^0=y^0$.

The non-linear $\s$-model sector requires a little more work and
we here just quote the relevant results.  
The canonical momenta are
\ba
\Pt^A:= \frac{\d \cL}{\d P_0^A} = \r \big( \gt^{00} P_0^A 
    + \gt^{01} P_1^A \big) \;\;\; , \;\;\;
\phi^\a:= \frac{\d \cL}{\d Q_0^\a} = 0    \la{canmom2}
\ea
Clearly,
\be
\phi^\a (x) \approx 0
\ee
is the constraint implementing the local $H$ invariance of the theory
as these generators satisfy
\be
\{ \phi^\a(x), \phi^\b (y) \} =  {f^{\a \b}}_\g \phi^\g (x) \dxy  
\la{Dirac2}
\ee 
The remaining non-vanishing brackets are given by
\ba
\{ \Pt^A (x), \Pt^B (y) \} &=&  {f^{AB}}_\a \phi^\a (x) \dxy  \non
\{ \Pt^A (x), P_1^B (y) \} &=& {f^{AB}}_\a Q_1^\a (x) \dxy 
     - \d^{AB} \:\p_1 \dxy \non
\{ \Pt^A (x) , Q_1^\a (y)\} &=& {f^{A\a}}_B P_1^B (x) \dxy \la{Dirac3}
\ea
and
\ba
\{ \phi^\a(x), \Pt^A (y) \} &=&  {f^{\a A}}_B \Pt^B (x) \dxy  \non
\{ \phi^\a(x), P_1^A (y) \} &=& {f^{\a A}}_B P_1^B (x) \dxy \non 
\{ \phi^\a(x), Q_1^\b (y)\} &=& {f^{\a\b}}_\g Q_1^\g (x) \dxy \la{Dirac4}
\ea

With the following parametrization of the worldsheet metric
in terms of lapse $N^0$ and shift $N^1$ 
\be
{e_\m}^{\a}= \pmatrix{N^0 & N^1 \cr 0 & {e_1}^{1} \cr}
  \;\;  \Longrightarrow  \;\;
\gt_{\m\nu} = \pmatrix{ N^0- \frac{(N^1)^2}{N^0} & -\frac{N^1}{N^0} \cr
          -\frac{N^1}{N^0} &  -\frac{1}{N^0}\cr}
\ee
(which differs slightly from the usual one because $\gt_{\m\nu}$ is 
unimodular), a straightforward calculation leads to the expected result
\ci{Dira67,HenTei92}
\be
\cL = \Pi \p_0 \r + \Pit \p_0 \rt + P_0^A \Pt^A - H[N^0,N^1,Q_0^\a]
\ee
Here 
\be 
H[N^0,N^1,Q_0^\a]:= \int dx \Big( N^0(x) \H(x)+ N^1(x)\cP(x) -
    Q_0^\a (x) \phi^\a (x) \Big)
\ee
is the total Hamiltonian with the Hamiltonian (WDW) constraint
\ba
\H (x):= \Pi \p_1 \rt + \Pit \p_1 \r + \ft12 \r^{-1} \Pt^A \Pt^A
         + \ft12 \r P_1^A P_1^A \approx 0
\ea
and the (spatial) diffeomorphism constraint
\be
\cP (x) := \Pi \p_1 \r + \Pit \p_1 \rt +  \Pt^A P_1^A \approx 0
\ee
In the remainder we will mostly work with the combinations
\ba
T_{\pm \pm} (x) = \ft12 \big(\H(x) \pm \cP(x)\big)  
    &=& \pm \Pi_\pm \p_1 \r^\pm + \r P^A_\pm P^A_\pm \non
    &=& 2\Pi_\pm \p_\pm \r +  \r P^A_\pm P^A_\pm 
\la{Vir1}
\ea
where
\be
\Pi_\pm := \ft12 ( \Pi \pm \Pit )
\ee
are canonically conjugate to $\r^\pm$. Obviously, $T_{\pm\pm}$
generate infinitesimal conformal diffeomorphisms $x^\pm\rightarrow
\tilde x^\pm (x^\pm)$. The constraints $T_{\pm\pm}(x)\approx 0 $ are
the analogs in our model of the Virasoro constraints of string theory;
besides, they are just the first order equations of motion
\Ref{eqmconf} for the conformal factor.  Note also that by Weyl
invariance of \Ref{Lagr} we automatically have $T_{+-}(x)\equiv
0$. With the above commutation relations it is straightforward to
check that \Ref{Vir1} generate two commuting classical Virasoro
algebras (i.e., without central term).

To fix the gauge we identify the dilaton $\r$ and the axion $\rt$ with
the world-sheet coordinates $(x^0,x^1)$ in the given coordinate
patch. For axisymmetric stationary gravity this is a well known trick
\ci{KSHM80} leading to Weyl canonical coordinates (which usually are
assumed to be global coordinates in the upper half plane, although
this is not necessary \ci{KorNic94}). For Lorentzian world-sheets,
there is an extra subtlety because we must distinguish whether the
vector $\p_\m \r$ is spacelike, timelike or null; these cases
correspond to physically distinct situations.\footnote{Yet another
inequivalent choice, appropriate for colliding plane wave solutions
which we will not discuss here, is $\r(x)= 1- (x^+)^2 - (x^-)^2$.} In
fact, there are cosmological models \ci{Gowd74} where the
signature $\p_\m\r\p^\m \r$ varies over different regions, thus
supporting a canonical treatment according to our local point of view.
For definiteness we will from now on assume that
\be
 G[f]:=\int dx f(x)\p_1 \r (x)= 0 \;\;\; , \;\;\;
 \tilde G[f] := \int dx f(x) (\p_1 \rt (x) -1) =0 \la{gaugefix1}
\ee
for {\em all} smooth functions $f\in C^\infty_0({\rm I})$ with compact
support in the given coordinate patch $\rm I$ (so that boundary terms
arising in partial integration can be dropped). Hence, by the
equations of motion \Ref{eqdil}
\be
\r (x)= x^0 + \r_0 \;\;\; , \;\;\; \rt (x) = x^1 + \rt_0
\la{gaugefix2}
\ee
since the zero modes $(\r_0,\rt_0)$ cannot be gauge fixed.
Thus the dilaton $\r>0$ serves as a ``clock field" (i.e., time is
measured by the size of the internal Kaluza Klein universe
which has an initial singularity at $\r=0$), while the axion 
$\rt$ would have to be interpreted as a ``measuring rod 
field"\footnote{Or, as H.~Weyl would have called it, 
``Ma\ss stabfeld".}.

Identifying the coordinates with $\r$ and $\rt$ in this way amounts to
solving the constraints $\H(x) = 0 \, , \, \cP (x) = 0$ up to their
zero (constant) modes. The remaining constraints $H =\int \H dx$ and
$P=\int \cP dx$ then generate the true dynamics of our model; in the
quantum theory they will effectively lead to two ``Schr\"odinger
equations" in the two-time formalism. To check the admissibility of
the gauge choice \Ref{gaugefix1}, we note that
\ba
\{ H[N^0,N^1,Q_0^\a] , G[f] \} &=& \int dx f(x)\p_1 N^0(x)  \non
\{ H[N^0,N^1,Q_0^\a] , \tilde G[f]  \} &=& \int dx f(x) 
\p_1 N^1(x) \la{gaugefix3}
\ea
For any non-zero $(N^0,N^1)\in C^\infty_0 ({\rm I})$, this indeed
cannot vanish for all $f\in C^\infty_0$. Therefore such $(N^0,N^1)$
generate ``bubble-like" deformations which we consider as pure gauge.
On the other hand, the constraints $H$ and $P$ correspond to
{\em constant} (non-zero) lapse and shift, and one easily sees that 
for them, the r.h.s.~of \Ref{gaugefix3} vanishes. This means that
the above gauge choice is preserved by the evolution generated
by $H$ and $P$.  
In analogy with \Ref{Vir1}, we will now consider the
mutually commuting constraint generators  
\be
C_\pm := \ft12 (H\pm P)
\ee
(for closed strings, we would have $C_+ \equiv L_0 \, ,\, C_-\equiv \bar L_0$.)
These light-cone Hamiltonians generate the dynamics along $x^\pm$ 
in the local patch, and form the basis of the ``two-time formalism". 
It is straightforward to check that commutation with 
$C_+ + C_-$ reproduces the equations of motion: for instance, we get
\be
\p_0 \Pit = \p_1 \Pi
\ee
which is consistent with \Ref{Pisigma}, and 
\be
\p_0 \Pi = \p_1 \Pit + \ft12 \r^{-2} \Pt^A \Pt^A -
    \ft12 P_1^A P_1^A
\ee
which is just the second order equation \Ref{eqmconf2} for the
conformal factor.

A proper analysis would require that we treat the constraints $\H(x)=
0$ and $\cP (x)= 0$ together with the gauge fixing conditions
\Ref{gaugefix1} as an infinite system of second class constraints,
with only the zero mode parts of $H$ and $P$ remaining as first class
constraints (since only for them, the r.h.s.~of \Ref{gaugefix3}
vanishes). This procedure shrinks the full phase space of the
covariant theory down to the so-called {\em reduced phase space},
which we will be mainly concerned with in the sequel. It not only
eliminates the $\d$-functions in the canonical brackets but also leads
to a replacement of the constraint functionals by ordinary functions.
In particular, the zero mode constraints from \Ref{Vir1}
become\footnote{For spacelike $\p_\m\r$ we would have $\cC_\pm= \pm 
\Pi_\pm + \r P_\pm^A P_\pm^A$.}
\be
\cC_\pm= \Pi_\pm + \r P_\pm^A P_\pm^A  \approx 0 \la{Vir2}
\ee
where we have replaced the integrated constraints $C_\pm$
by the corresponding densities $\cC_\pm$ at an arbitrary
but fixed space-time reference point in the patch. This is permitted
because after solving the non-zero mode part of the constraints, only
the constant part of the densities remains, and consequently $C_\pm$ 
and $\cC_\pm$ differ merely by an irrelevant volume (length) factor.
Furthermore, we now have the reduced phase space brackets
\be
\{ \Pi_+ , \r^+ \} = \{ \Pi_- , \r^- \} = 1  \;\;\; , \;\;\;
\{ \Pi_+ , \r^- \} = \{ \Pi_- , \r^+ \} = 0 \la{Dirac5}
\ee
where again all variables are to be taken at the chosen fixed
reference point in space-time. These brackets result from \Ref{Dirac1}
after removal of the non-zero mode contributions to the $\d$-function.

Upon quantization, \Ref{Vir2} will become ordinary differential
operators rather than functional differential operators unlike the 
original constraints \Ref{Vir1}; similarly, the WDW functional of the
covariantly quantized theory will become a (Hilbert-space-valued)
function of the coordinates $x^\pm$. We will return to this
in chapter 5. However, we must first discuss the role 
of \Ref{Vir2} for the matter variables.

\subsection{Poisson Structure and Hamiltonians for the Matter Sector}

For the matter sector, an {\em ab initio} derivation of the new
Poisson structure is not yet available, and we will thus simply
postulate the brackets in such a way that the correct equations of
motion are obtained. The new Poisson structure presented here amounts
to a {\em de facto} resolution of certain technical problems of the
corresponding flat space models (for instance having to do with the
non-ultralocal term on the r.h.s.~of \Ref{Dirac3} \ci{VeEiMa84}),
mainly because spatial $\d$-functions and their derivatives are
altogether absent in our formalism. Before writing the brackets down,
however, we give the light-cone Hamiltonians\footnote{By a slight
abuse of language, we will sometimes refer to the matter part of
\Ref{Vir2} simply as the (matter) ``Hamiltonians'',
cf.~\Ref{Hamiltonian}.} in the Weyl gauge \Ref{gaugefix1}. They are
\ba
H_\pm & := & \r^{-1}
\sum_{k,l}\frac{\tr(B_kB_l)}{(1\pm\g_k)(1\pm\g_l)}
- \r^{-1} \sum_{k,l}\frac{\tr (B_kB_l)}{(1\pm\g_l)} \non
&=& \r^{-1} ~\tr \left(B(\mp1)\right)^2 
- \tr (B_\i P_\pm)    \la{H} 
\ea
The first term on the r.h.s.~can be directly deduced from \Ref{Vir2}
by substitution of \Ref{rel} into the constraints \Ref{Vir2}, 
remembering that $\p_\pm \r = \ft12$ in this gauge. The second term
involving $B_\i$ cannot be motivated in this way. As we will see
in a moment, however, it vanishes once the coset constraint
$B_\i =0$, which is not automatically included in the 
deformation equations unlike in \Ref{Vir2}, is properly taken 
into account. With this caveat, we can claim that the constraints
\be
\cC_\pm = \Pi_\pm + H_\pm \approx 0   \la{Cpm}
\ee
indeed coincide with \Ref{Vir2}.

The Poisson brackets with the requisite properties for the spectral 
parameter current $B(\g)\!\equiv\!B^a(\g) Z_a$ turn out to be 
\be\la{hbracket}
\{B^a(\g) \, , \, B^b(\g')\} 
     = -{f^{ab}}_c \, \frac{B^c(\g)-B^c(\g')}{\g-\g'}
\ee
where the space-time coordinate is kept fixed (hence ``equal point 
brackets"). These brackets may be written in several equivalent ways.
Depending on their background readers might prefer the 
explicit index notation
\be
\{B_{\a\b}(\g)\, , \, B_{\gamma\d}(\g')\} = \frac1{\g-\g'}
\Big(\d_{\gamma\b}[B(\g)-B(\g')]_{\a\d} - \d_{\a\d}
[B(\g)-B(\g')]_{\gamma\b}\Big)
\ee
for the matrices $B_{\a\b}(\g)$, or the compact 
tensor notation of \ci{Fadd84}
\be\la{hbtens}
\{B(\g) \stackrel{\otimes}{,} B(\g') \} = 
\left[B(\g)\otimes I + I\otimes B(\g'),\frac{\Omega}{\g-\g'}\right]
\ee
with the Casimir operator $\Omega\!\equiv\!Z_a\!\otimes\! Z^a$.  We
emphasize that the Poisson structure \Ref{hbracket}, which is
ubiquitous in integrable systems \ci{Fadd84}, cannot be ``amended"
by the introduction of spatial $\d$-functions, as otherwise it would
be incompatible with the deformation equations \Ref{compB}. On the 
contrary, these equations {\em define} how to continue the bracket to
other space-time points.

The brackets can be alternatively expressed in terms of the
singularities and residues of the spectral parameter current. We leave
it as an exercise to show that \Ref{hbracket} in the parametrization
of \Ref{spcurrent} is equivalent to:
\be\la{resbracket}
\{B_j^a,B_k^b\} = \d_{jk} {f^{ab}}_c   B_k^c, \qquad
\{B_j^a,\g_k\} = 0, \qquad
\{\g_j,\g_k\} = 0
\ee

We now have all the tools ready to check that that the
deformation equations \Ref{defB} admit a Hamiltonian formulation 
with respect to the Poisson structure \Ref{hbracket} 
and the Hamiltonians \Ref{Cpm}. Indeed, 
\ba
&& \left\{\tr \left(B(\mp1)\right)^2 ,~ B(\g)\right\} ~=~ 
2\;\frac{[B(\mp1),B(\g)]}{\g\pm1} \non
&\Longrightarrow& \left\{\tr \left(B(\mp1)\right)^2 ,~ B_j~\right\} 
~=~ 2\,\sum_{k} \frac{[B_j,B_k]}{(1\pm\g_j)(1\pm\g_k)} 
\ea
and 
\be
\left\{\tr (B_\i P_\pm),B_j\right\} = [B_j,P_\pm]
\ee
In summary, we have formulated the model as a Hamiltonian system 
such that the deformation equations can be obtained from
\be\la{Hamiltonian}
D_\pm B_j = \{H_\pm,B_j\}
\ee
The compatibility of the deformation equations that was already
emphasized above is equivalently expressed by the fact that the
Hamiltonians $H_\pm$ have vanishing mutual Poisson bracket;
consequently, the corresponding flows commute. However, we still have
to take care of the coset constraints of section \ref{Sconstraints}.
In a Hamiltonian formulation they must be properly treated \`a la
Dirac \ci{Dira67,HenTei92}. We briefly describe the result of this
procedure. The nontrivial constraint is \Ref{coset}.  The detailed
analysis shows that only
\be
B_\infty + \n(B_\infty) \approx 0 \la{1class}
\ee
remains as a first-class constraint, whereas all other constraints
(among them in particular $B_\i- \n(B_\i)$) allow explicit resolution,
such that after the Dirac procedure they vanish strongly.  The Poisson
structure of $B(\g)$ is modified to \ci{KorSam96}
\ba\la{DiracB}
\{B^a(\g),B^b(\g')\}_{DB} &=& 
-\f12 {f^{ab}}_c \, \frac{B^c(\g)-B^c(\g')}{\g-\g'}\\ 
&&{}+\f12 {f^{a\n(b)}}_c \, \frac{B^c(\g)}{\g'-\f1{\g}}
+\f12 {f^{\n(a)b}}_c  \, \frac{B^c(\g')}{\g-\f1{\g'}},\nn
\ea
with the notational convention (and choice of basis)
$Z^{\n(a)}\!\equiv\!\n(Z^a)$.

In terms of the residues $B_j$ this implies (if we for simplicity
restrict to the case \Ref{polreal}):
\be
\{B_j^a,B_k^b\} = \ft12\d_{jk}f^{abc}B^c_k\qquad
\mbox{for $j,k\le n$}
\ee
as well as the strong identities
\be
B_j = \n(B_{j+n}),
\ee
For $B_\i$ this already implies, that
\ben
B_\i=\sum_{j=1}^{n} \Big(B_j + \n(B_j)\Big) \in {\mathfrak h},
\een 
whereas the remaining constraint 
\be
\sum_{j=1}^{n} \Big(B_j + \n(B_j)\Big) = 0
\ee
remains first class according to \Ref{1class}.
Thus, the Poisson structure is defined on half of the variables $B_j$,
whereas for the other half it is determined by the solution of the
constraints. After the Dirac procedure, the second term of the
Hamiltonian \Ref{H} may be dropped, since the relations
$B_\i=\n(B_\i)\!\in\!\mathfrak{h}$ and $P_\pm\!\in\!\mathfrak{k}$ hold
strongly.
\bigskip

Finally, we would like to study the action of
the surviving first class constraint \Ref{1class}. It is a simple
exercise to show that this expression generates the gauge
transformations
\be
B(x;\g)\mapsto {\rm h}^{-1}(x)B(x;\g){\rm h}(x)\qquad\mbox{with}
\quad{\rm h}(x) \in H
\ee
Owing to \Ref{rel}, these are precisely the $H$-valued gauge
transformations \Ref{gauge} of the original theory. 
The Hamiltonians 
\be\la{HD}
H_\pm = \r^{-1}~\tr B^2(\mp1)
\ee
generate deformation only modulo $H$ gauge transformations (as evident
from the appearance of the covariant derivative in
\Ref{Hamiltonian}). Thus, the first class constraint \Ref{1class} can
be added with impunity and moreover be employed to keep the dynamics
in a fixed section of the gauge orbits.

\subsection{Hamiltonian Formalism for Non-Isomonodromic Configurations}
As explained in section \ref{beyiso}, the natural generalization of
the residues $B_j(x)$ for non-isomonodromic solutions is the density
$\B(x;w)$, $w\in \ell$.  The deformation equations \Ref{defB} are
replaced by their continuous analogs \Ref{DEFB}, such that the
isomonodromic ansatz can be recovered in a special case \Ref{Biso}.
Inspection of \Ref{Biso} immediately suggests a Poisson structure for
$\B(w)$: in the isomonodromic sector we can rewrite \Ref{resbracket}
as\footnote{We suppress the $x$-dependence from now on since all
brackets are to be taken at a fixed reference point.}
\ba
\{\B^a(w),\;\B^b(v)\} &=& \left\{\sum_{j=1}^{N}B^a_j\delta(w-w_j)\;,\;
\sum_{j=1}^{N}B^b_j\delta(v-w_j)\right\}\non
&=&\sum_{j=1}^{N}{f^{ab}}_c B_j^c\delta(w-w_j)\delta(v-w_j)
\non
&=&\sum_{j=1}^{N}{f^{ab}}_c B_j^c\delta(w-w_j)\delta(w-v)
~=~-{f^{ab}}_c  \B^c(w)\delta(w-v)\nn
\ea
Regarding $\B(w)$ as the basic variables, we are thus led to postulate
\be
\{\B^a(w),\;\B^b(v)\}=-{f^{ab}}_c \;\B^c(w)\delta(w-v)\;\;\;\;\;\;
v,w\in \ell
\la{PBBc}\ee
which has precisely the structure of an affine Lie algebra.
The bracket for the spectral parameter currents $B(\g(w))$ is then
a direct consequence of the representation \Ref{Cau}:
\ba
\{B^a(\g(w)),\;B^b(\g(v))\}&=&
\left\{\oint_{\p D} \f{\B^a(w'){\rm d}w'}{\g(w')-\g(w)}\;,
\oint_{\p D} \f{\B^b(v'){\rm d}v'}{\g(v')-\g(v)}\right\}\non
&=&\oint_{\p D}\oint_{\p D}
\f{\{\B^a(w'),\;\B^b(v')\}d w' dv'}{(\g(w')-\g(w))
(\g(v')-\g(v))}\non
&=& -{f^{ab}}_c \oint_{\p D}\f{\B^c(w')d w'}{(\g(w')-\g(w))
(\g(w')-\g(v))} \non
&=&
 - {f^{ab}}_c  \;\f{B^c(\g(w))-B^c(\g(v))}{\g(w)-\g(v)}\nn
\ea
which indeed coincides with \Ref{hbracket}. The Dirac bracket
\Ref{DiracB} would follow in the same fashion after the Dirac
procedure of \Ref{cB} at the level of $\B(w)$. 

An obvious question at this point is whether we can include a central 
term in the bracket \Ref{PBBc}, i.e.~replace \Ref{PBBc} by
\be
\{\B^a(w),\;\B^b(v)\}=-{f^{ab}}_c\; \B^c(w)\delta(w-v)
    + K \n^{ab} \p_w \d (w-v) \;\;\;\;\;\; v,w\in \ell
\la{PBBc1}\ee
Remarkably, this is indeed possible because the holomorphic bracket
\Ref{hbracket} is insensitive to this modification, as the central
term drops out in the above integral.  It is not clear at the moment,
whether this central extension is related to the central term
appearing in the Geroch algebra \ci{BJ2,BreMai87, Nico91}.
\bigskip

The matter Hamiltonians $H_\pm$ in \Ref{H} are re-expressed as follows:
\be
H_{\pm} = 
 \r^{-1}\p_\pm\rho~\tr\left[\oint_{\ell\cup \ell^*} 
\f{\B(w){\rm d}w}{1\pm\g(w)}\right]^2 
+ \tr \left[\oint_{\ell\cup \ell^*} \B(w)d w P_\pm\right] ,
\la{HpmB}\ee
which shows that in the general scheme they are non-local in the
spectral parameter plane. This is one indication how the non-locality
in space-time is converted into non-locality in the spectral parameter
plane in this formalism.

The Dirac procedure will again kill the second term of the
Hamiltonian. Solving the coset constraint \Ref{cB} here amounts to
defining the Poisson structure on one sheet of the Riemann surface and
transporting it to the other sheet by means of the constraint. The
surviving first class constraint \Ref{1class} takes the form
\be
\B_\i\equiv\oint_{\ell\cup \ell^*} \B(w)d w \equiv \oint_\ell  
\Big(\B(w)+\n(\B(w^*))\Big)dw =0
\ee

{}From \Ref{HpmB}, \Ref{PBBc} and \Ref{DEFB} we see that the total
dynamics of $\B(w), w\!\in\!\ell$ is generated by the matter
Hamiltonians $H_\pm$. The variables $B((\g(w))$ on the contrary also
carry an explicit $x^\pm$-dependence via $\g$, whose dynamics is not
generated by these Hamiltonians. This is in complete analogy with the
isomonodromic sector, where an explicit $x^\pm$-dependence solely
originated from the $x$-dependence of the locations of the poles
$\g_j$.

\subsection{Conserved Non-Local Charges}

We have already encountered some conserved quantities
in the isomonodromic sector. In this section, we shall construct an
even larger set of observables that is complete in the sense
that the solution in any isomonodromic sector with a fixed number of
poles may (generically) be uniquely reconstructed from
these data . With the additional structure of Poisson brackets at
our disposal, we can proceed to study the algebraic structure 
of these observables.

Returning to the deformation equations \Ref{defB}, it is obvious, that
the set of eigenvalues of the residue matrices $B_j$, encoded into the
traces $\tr B_j, ~\tr B_j^2, \dots$ is independent of the coordinates
$x^\pm$. However, according to \Ref{hbracket} the Poisson structure of
these eigenvalues turns out not to be too interesting because they all
commute. This means that we have to fix all these quantities by hand
in order to achieve non-degeneracy of the Poisson structure. A more
``advanced'' construct regarding the isomonodromic ansatz
\Ref{spcurrent} are the monodromy matrices related to this current,
which are defined by
\be
\vh(\g)\mapsto M_j\vh(\g) \qquad\mbox{for $\g$ encircling $\g_j$}
\ee
These monodromies indeed encapsulate the complete information about the
spectral parameter current \Ref{spcurrent} itself (after resolution of
the classical Riemann-Hilbert problem). As they are obviously
$x^\pm$-independent, they build a promising object for observables of
the model.

Their mutual Poisson brackets calculated from \Ref{hbracket} exhibit
the following qua\-dra\-tic structure \ci{KorSam96}
\ba
\{M_i^1, M_i^2\} &=& 
i\pi\,\Big( M_i^2\,\O \,M_i^1 - \,M_i^1\,\O \,M_i^2
\Big)\label{monoMiMi}\\
\{M_i^1,M_j^2\} &=& 
i\pi \,\Big( M_i^1\,\O\,M_j^2 + M_j^2\,\O\,M_i^1  -
\O\,M_i^1M_j^2 - M_i^1M_j^2\,\O\Big),\label{monoMiMj}\\ 
&&{\mbox {for}}\enspace i<j\nn
\ea
where we have introduced the shorthand notation $M_i^1\equiv
M_i\otimes I$ and $M_i^2\equiv I\otimes M_i$.  This algebra and its
quantization have been extensively discussed in the lectures of
Alekseev at this School \ci{AlekC96}.  We will therefore only comment
on some of its features to our model.
\begin{itemize}
\item{The algebra of monodromy matrices includes the first-class
constraint
\be
M_\i\equiv \prod_{i=1}^N M_i = I, \label{conM}
\ee 
which generates common conjugation. This is the analogue of
\Ref{1class} in terms of the spectral parameter current. Gauge
invariant objects are built from traces of arbitrary products of
monodromy matrices.}

\item{The precise definition of the monodromy matrices depends on the
normalization $\vh \big|_{\g=\i} = \n(\cV)$ (cf.~\Ref{BZ}) ---
otherwise they are defined only up to gauge transformation. The
distinguished path $[\g\!\rightarrow\!\i]$ in the $\g$-plane gives
rise to a cyclic ordering of the monodromy matrices, that defines
\Ref{monoMiMj} and also \Ref{conM}. It is a remnant of the so-called
eyelash that enters the definition of the analogous Poisson structure
in the combinatorial approach \cite{FocRos92,AlGrSc95a,AlekC96}, being
attached to every vertex and representing part of the freedom in this
definition.}

\item{An apparent obstacle of the structure (\ref{monoMiMi}),
(\ref{monoMiMj}) is the violation of Jacobi identities. Actually, this
can be traced back to the constraint (\ref{1class}) in the
calculation of the Poisson brackets. As these brackets are 
valid only on the first-class constraint surface \Ref{conM}, Jacobi
identities cannot be expected to hold in general.

However, the same reasoning shows, that the structure
(\ref{monoMiMi}), (\ref{monoMiMj}) restricts to a Poisson structure
fulfilling the Jacobi identities on the space of gauge invariant
objects. On this space, the structure coincides with the restrictions
of previously found and studied structures on the monodromy matrices
\cite{FocRos92,AlekC96}:
\ba
\{M_i^1 \, ,\, M_i^2\} &=& 
M_i^2r_+M_i^1 + M_i^1r_-M_i^2 -
r_-M_i^1M_i^2-M_i^1M_i^2r_+ 
 \non
\{M_i^1 \, , \,  M_j^2\} &=& 
M_i^1r_+ M_j^2 + M_j^2r_+ M_i^1 -
r_+M_i^1M_j^2 - M_i^1M_j^2r_+   \non
 &&  {\rm for }\enspace i<j, \la{FRMB} 
\ea
where $r_+$ and $r_-\!:=\!-\Pi r_+ \Pi$ are arbitrary solutions of
the classical Yang-Baxter equation
\be\label{cYB}
[r^{12},r^{23}] + [r^{12},r^{13}] + [r^{13},r^{23}] = 0.
\ee
{}From the quantum point of view, non-associativity of \Ref{monoMiMi},
\Ref{monoMiMj} may be understood as a classical limit of the associated
quasi-quantum groups \ci{KorSam96}.}
\item{The previously found conserved quantities $\tr B_j, ~\tr B_j^2,
\dots$ are embedded into the algebra of monodromy matrices via 
\ben
M_j \sim e^{2\pi i B_j}, 
\een 
their trivial brackets corresponding to the trivial brackets of
eigenvalues of monodromy matrices according to \Ref{monoMiMi},
\Ref{monoMiMj}.  }

\item{In fact, the structure \Ref{monoMiMi}, \Ref{monoMiMj} has been
calculated from the Poisson structure \Ref{hbracket}; the coset
constraints \Ref{coset} (except the first-class part \Ref{1class})
have not yet been taken into account. Thus, an additional Dirac
procedure is also required on the level of this algebra. The coset
constraints in terms of monodromies translate into existence of a
matrix $C_0$ with
\ben
M_{j} = C_0\n(M_{j+2n})C_0^{-1}
\een
The space of gauge invariant objects therefore admits a decomposition
\ben
M_S \oplus M_{A}
\een
according to the involution $\n^\i$ defined by $\n^{\i}(M_j)= C_0
\n(M_{j+2n}) C_0^{-1}$. This involution is an automorphism of
\Ref{monoMiMi}, \Ref{monoMiMj}, such that its invariant subspace $M_S$
forms a closed subalgebra invariant also w.r.t.~the Dirac procedure.
}
\end{itemize}
\bigskip

Abandoning our local point of view for the moment, let us briefly
describe the relation of these observables found in the
isomonodromic sector to the conserved charges that have been identified
within the conventional formulation of the models. A simple consequence
of the linear system \Ref{ls} is that for asymptotically flat solutions 
the function $\vh$ becomes constant at spatial infinity:
\be
\vh(w,x^0,x^1\!\rightarrow\!\i) \rightarrow \vh_\i(w)
\ee
Its monodromies --- for $\g$ encircling $\g_j$ or equivalently $w$
encircling $w_j$ --- that we have discussed above, certainly survive
this limit. We can now translate our quantities back into the 
Breitenlohner-Maison picture (cf.~\Ref{BZ}) by
$\vh_{BM}\!=\!S(w)\vh_{BZ}$. Making use of a result of \ci{BreMai87}
according to which this function tends to unity at spatial infinity,
we arrive at the identification
\be
\vh_\i(w) = S(w)^{-1} 
\ee

Thus, we have essentially succeeded in computing the algebra of the
monodromies of the matrix $\cM_{BM}(w)$, which plays the pivotal
role in the approach of \ci{BreMai87}.
This suggests yet another route for comparing our Poisson structure
to the canonical one: calculation of the canonical Poisson bracket
between these conserved charges should lead to the same result. 
Unfortunately, the conventional approach got stuck 
before attaining this aim.

\section{QUANTIZATION}
\setcounter{equation}{0}
\subsection{General Remarks}

The quantization of the models will be based on the reduced
phase space description corresponding to the Weyl gauge (alias
the light-cone gauge) \Ref{gaugefix1} introduced in section 4.1. 
The WDW equation and diffeomorphism constraint of the covariant 
approach, which are functional differential equations, are thereby
reduced to the partial differential equations \Ref{physstate2} below,
which we will simply refer to as the ``quantum constraints", and
which encode the dynamical content of the model.
There is, of course, the question whether by quantizing on the reduced 
phase space, possible anomalies of the constraint algebra of
$2d$ diffeomorphisms (i.e.~the Virasoro algebra) might have been swept 
under the carpet. However, just as in light-cone gauge string theory, 
such anomalies are not directly visible in this gauge, 
but would manifest themselves in some analog of the target 
space Lorentz algebra (whose identification in the present context
would amount to a ``stringy" interpretation
of these models \ci{KorNic96}).

Accordingly, we perform the textbook substitution
\be
\{\, .\, , \, . \, \}\longrightarrow \f{1}{i\hbar} [\, . \, , \, . \,]
\ee
in all canonical brackets, converting all phase space variables into 
operators. Since $\r^\pm=x^\pm + \r_0^\pm$ in this gauge, we set
\be
\Pi_\pm = i\hbar \p_\pm 
\ee
The constraints \Ref{Vir2} are thereby transmuted into differential
operator constraints acting on the quantum Hilbert space of the theory.
The quantum states $\Phi\equiv\Phi (x^\pm)$ are
space-time coordinate dependent elements of a Hilbert space
$\H$ to be specified below whose precise structure depends on
the matter sector. {\em Physical states} are by definition those
elements of $\H$ which are annihilated by the constraint operators, i.e.
\be
\Phi(x)\in\H_{phys} \Longleftrightarrow
\cC_\pm (x)\Phi(x) = 0  \la{physstate1}
\ee
In other words, physical states must satisfy the differential equations
\be
\big( i\hbar \p_\pm + H_\pm) \Phi (x) = 0 \la{physstate2}
\ee 
where we must now insert the Hamiltonians from \Ref{H}.
In addition, physical states may be subject to further constraints
related to the local internal symmetries of the model. The two equations 
\Ref{physstate2} resemble the time-dependent Schr\"odinger equation, as
the matter Hamiltonians explicitly depend on the coordinates $x^\pm$.
We stress that the operators on the l.h.s.~of \Ref{physstate2} 
are to be interpreted as total derivatives $d/dx^\pm$. In other words,
\Ref{physstate2} is nothing but the statement that the solutions $\Phi$
should not depend of the coordinates; we thus have a rather
simple realization of the idea that physical states in quantum
gravity should be invariant under the full set of $2d$ coordinate
transformations!

The difficult part in solving these equations is the analysis
of the matter sector governed by the Hamiltonians $H_\pm$,
which are non-trivial operators acting in the Hilbert space $\H$. 
By contrast, for the Einstein-Rosen waves and for the (closed) string
the operators $H_\pm$ are free field Hamiltonians, and the equations 
can be dealt with by standard Fock space methods; e.g. for the 
closed string, $\H_{phys}$ would just be the Fock space 
of two sets of transverse oscillators, and the equations \Ref{physstate2} 
would be solved by demanding $L_0= \bar L_0$ and 
$M^2=\sum_{n=1}^\i \a^i_{-n}\a^i_n+\sum_{n=1}^\i \bar\a^i_{-n}\bar\a^i_n-2$
on the physical states. This path is obstructed here, because
even in the flat space limit, the non-linear $\s$-model can at best
be treated perturbatively with these methods. On the other hand,
the methods borrowed from the theory of integrable systems, which
we employ here, are essentially non-perturbative. 
Our results furthermore indicate that the underlying group 
theoretical structure will play an essential role in the further
development of the subject. The rest of this section is therefore
devoted to a discussion of the matter sector. As we will explicitly
demonstrate, the central equations \Ref{physstate2} can be reduced 
to the KZ equations. The recourse to
techniques borrowed from conformal field theory thus leads to a
substantial simplification of \Ref{physstate2}.  

\subsection{Canonical Quantization of the Matter Sector}
As we explained before, the matter sector can be equivalently described
in terms of $B(\g)$ or $\B(w)$, but for clarity of presentation
we will restrict attention to the isomonodromic sectors (hence
``isomonodromic quantization"). To this aim, let us
immediately proceed to quantize the Poisson bracket \Ref{hbracket}. 
We arrive at the following commutation relations
\be
[B^a(\g),B^b(\g')] = 
-i \hbar {f^{ab}}_c \, \frac{B^c(\g)-B^c(\g')}{\g-\g'}
\la{CR10a}
\ee
or, equivalently,
\be\la{CR20}
[B_j^a,B_k^b] =  i\hbar \d_{jk}{f^{ab}}_c B_k^c 
\ee
(remember that we suppress the $x$-dependence in these equations).
The spectral parameter current and its residues have thus
become operators on a Hilbert space, which for a given isomonodromic
(``$N$-wave") sector is the direct product
\be
\H=\cH_1\otimes...\otimes \cH_N
\ee
of $N$ representation spaces of the Lie algebra ${\mathfrak g}$
or its complexification ${\mathfrak g}_\C$, one for each set of
operators $B_j \equiv B_j^a Z_a$. Note that every $B_j$ is a matrix, 
all of whose entries are operators on $\H_j$.\footnote{This is the
poor man's definition of a quantum group.} For obvious reasons
we shall assume the spaces $\H_j$ to be {\em unitary representation spaces},
since this will automatically lead to a positive definite scalar
product in the Hilbert space associated with the reduced phase space
quantization. As is well known the structure of unitary irreducible 
representations of the non-compact group $G$ and its complexification $G_\C$ 
are very different. Moreover, if we are dealing with a given real form
of a non-compact group, highest or lowest weight representations 
may or may not exist, depending on whether or not $H$ contains a 
$U(1)$ factor \ci{Murat,GunSac82}, with consequences for the spectrum
of physical parameters.

To illustrate these remarks we will mostly restrict attention in the
remainder to the group $G=SL(2,\R)$, whose representation
theory is well understood \ci{Murat}. Specializing the 
(isomonodromic) spectral parameter current to this case, we have
\be
B(\g)= \frac{i\hbar}{2}\pmatrix{ \hb(\g)& 2\eb (\g) \cr  
                                2\fb(\g)& -\hb(\g) \cr} 
  \;  \quad\Longleftrightarrow\quad \;
B_j = \frac{i\hbar}{2}\pmatrix{ \hb_j& 2\eb_j \cr  
                                2\fb_j& -\hb_j \cr} 
\ee
with
\be
\hb (\g) = \sum_{j=1}^N \frac{\hb_j}{\g - \g_j} \;\;\; , \;\;\;
\eb (\g) = \sum_{j=1}^N \frac{\eb_j}{\g - \g_j} \;\;\; , \;\;\;
\fb (\g) = \sum_{j=1}^N \frac{\fb_j}{\g - \g_j}     \la{Chev1}
\ee
The reality condition \Ref{reality} translates into
\be
\hb(\g)^\dagger = -\hb(\overline{\g}) \;\;\; , \;\;\;
\eb(\g)^\dagger = -\eb(\overline{\g}) \;\;\; , \;\;\;
\fb(\g)^\dagger = -\fb(\overline{\g}) 
\ee
For real poles (cf.~\Ref{polreal}), the entries of  
$B_j$ become hermitean, and the operators $(\hb_j,\eb_j,\fb_j)$
are just the anti-hermitean Chevalley generators of $N$ mutually 
commuting $SL(2,\R)$ groups, viz.
\be
[\hb_j,\eb_k]= 2\d_{jk} \eb_j \;\;\; , \;\;\;
[\hb_j,\fb_k]= -2\d_{jk} \fb_j \;\;\; , \;\;\;
[\eb_j,\fb_k]= \d_{jk} \hb_k
\ee
Hence, each $\H_j$ is a unitary representation space of $SL(2,\R)$.
For {\em complex} poles $\g_j(x)$ (cf.~\Ref{polrea} and \Ref{polcos}), 
on the other hand, the operators $(\hb_j,\eb_j,\fb_j)$ 
obey the same commutation relations as before, but
\be
\hb_j^\dagger = -\hb_{j+n} \;\;\; , \;\;\;
\eb_j^\dagger = -\eb_{j+n} \;\;\; , \;\;\;
\fb_j^\dagger = -\fb_{j+n}  \la{qcoset1}
\ee
In this case, we are dealing with $\ft12 N$ mutually commuting 
$SL(2,\C)$ groups. Readers may remember from the representation theory 
of the Lorentz group that $SL(2,\C)$ possesses no discrete unitary
representations at all. This is also evident from the fact that
${\mathfrak sl}(2,\C) \cong {\mathfrak so}(1,3)$ does not decompose 
into a sum of two mutually commuting subalgebras, unlike 
${\mathfrak so}(4)$ or ${\mathfrak so}(2,2)$. Consequently, there
is no escape from this difficulty. 

To analyze the coset constraints we will only consider real poles to
keep the discussion as simple as possible.  We recall that
$\n(X)=-X^t$ for $SL(2,\R)$; with the numbering of poles as in
\Ref{polreal}, we readily obtain
\be
\hb_j = - \hb_{j+n} \;\;\; , \;\;\;
\eb_j = - \fb_{j+n} \;\;\; , \;\;\; \fb_j = - \eb_{j+n} \la{qcoset2}
\ee
The condition \Ref{cosreal} reduces to
\be
\sum_{j=1}^n (\fb_j - \eb_j) = 0  \la{qcoset3}
\ee
In accord with our remarks in section 4.2, \Ref{qcoset2} should be regarded 
as second class constraints; indeed, \Ref{qcoset2} instructs us to 
eliminate all operators with index values $j>n$ in terms of the
remaining ones. On the other hand, \Ref{qcoset3} is first class, 
in agreement with \Ref{1class}: it is just the canonical generator 
of the $SO(2)$ gauge transformations on the states. Therefore, 
physical states by definition must be annihilated by \Ref{qcoset3}
and cannot carry any $H$ charge. 

For practical calculations it is oftentimes convenient to
switch to the $SU(1,1)$ Chevalley basis
\be
e_j:= \ft12 (-i\hb_j + \eb_j+\fb_j) \;\;\; , \;\;\;
f_j:= \ft12 (i\hb_j + \eb_j+\fb_j) \;\;\; , \;\;\;
h_j:= i(\fb_j - \eb_j)
\ee
The main advantage of this basis is that the relations 
$h_j^\dagger = h_j , e_j^\dagger = - f_j$ allow us to diagonalize 
the operator $h_j$ on the states and to interpret $e_j$ and $f_j$ 
as creation and annihilation operators. It should be emphasized
that these operators have nothing to do with the conventional Fock 
space creation and annihilation operators of free particles, 
as they create and annihilate ``collective excitations".
\Ref{qcoset2} now reads
\be
h_j=-h_{j+n} \;\;\; , \;\;\;
e_j=-e_{j+n} \;\;\; , \;\;\; f_j=-f_{j+n} \la{qcoset4} 
\ee
while \Ref{qcoset3} becomes
\be
\Big( \sum_{j=1}^n h_j \Big) \Phi = 0 \quad {\rm for} \quad
\Phi\in\H_{phys} \la{qcoset5}
\ee

\subsection{Quantum Constraints and KZ Equations}

We now return to section 3.5 where the classical constraints were
solved for the conformal factor in terms of the $\t$-function, and
show that this result has a precise quantum mechanical analog.  This
we do by reducing the quantum constraints \Ref{physstate2} to a
modified version of the KZ equations from conformal field theory
\ci{KniZam84}. In fact, disregarding the coset constraints \Ref{coset}
(i.e.~here \Ref{qcoset4}), we would arrive precisely at the KZ
equations, with the only difference that the worldsheet coordinates
$z_j$ labeling the insertions of conformal operators in the correlator
are replaced by the movable singularities $\g_j(x)$ in the spectral
parameter plane.  In this fashion one can see that the quantum analog
of the $\t$-function is just the physical state solving the quantum
constraints \Ref{physstate1}; hence, it is quite appropriate to call
$\Phi(x)$ the {\em quantum $\t$-function} (alternatively,\footnote{As
suggested to us by A.A.~Morosov.} one could reserve this name for the
quantum mechanical evolution operator, which is a ``matrix" whose
columns consist of an orthonormal basis in the space of solutions of
the KZ equations). However, due to the coset constraints, we obtain a
slightly modified version of the KZ-system, that we refer to as the
Coset-Knizhnik-Zamolodchikov (CKZ) system \ci{KorSam96}.  Since the
techniques for solving these modified equations have not yet been
elaborated, but might resemble the strategies followed in solving the
usual KZ equations \ci{SchVar90,Babu93}, we first describe the
quantization neglecting the coset constraints.
 
To prove the above assertions, we start from the ansatz
\be
\Phi(x) = F(x) \tP \big( \{ \g_j(x) \} \big)   \la{KZansatz1}
\ee
where $F(x)$ is an ordinary function and $\tP \in\H$ by assumption depends
on the coordinates only through the $\g_j(x)$. From \Ref{KZansatz1}
we get, using \Ref{despectral},
\be \la{KZansatz2}
\p_\pm \Phi = (\p_\pm F) \tP + \r^{-1}\p_\pm \r F
\sum_{j=1}^N \frac{\g_j(1\mp \g_j)}{1\pm \g_j} \frac{\p \tP}{\p \g_j} 
\ee
We next split this equation into two sets of equations, one for $F$
and one for $\tP$. To cut a long story short, we will assume 
the following equations to hold for $\tP$
\be
\frac{\p \tP}{\p \g_j} = \sum_{k\neq j}^N \frac{\Omega_{jk}}{\g_j - \g_k}\tP
\la{KZequation} \ee
which are just the famous KZ equations. To reconcile this ansatz with
the original equations \Ref{physstate2}, we must make the
identification
\be
\Omega_{jk} = \frac1{i\hbar}{\rm tr} \, B_j B_k
\ee
which e.g.~for $SU(1,1)$ leads to
\be
\Omega_{jk} = i\hbar \big( \ft12 h_j \otimes h_k +
         e_j \otimes f_k + f_j \otimes e_k \big)  \la{Om1}
\ee
Substituting the ansatz \Ref{KZansatz1} and \Ref{KZansatz2} into
\Ref{physstate2}, a little algebra shows that we can satisfy
the constraint, provided that
\be
 \left\{ i\hbar \p_\pm \log F + \r^{-1}\p_\pm \r {\rm tr} B_\i^2
 + \r^{-1} \p_\pm \r \sum_j {\rm tr} B_j^2
 \left( \frac{1}{(1\pm\g_j)^2} - \frac12 \right) \right\} \Phi = 0
\ee
This equation still contains the operators $\tr B_j^2 $ and $\tr
B_\i^2$, but it can be integrated in closed form if we assume that
they act diagonally on the quantum state $\Phi$.  For the known
solutions of the KZ equations, the validity of this assumption can be
verified by explicit computation \ci{KorNic96}. Designating the
respective eigenvalues by $(i\hbar)^2a_j$ and $(i\hbar)^2a_\i$
respectively, we arrive at the final result
\be
\Phi (x) = \r^{\f12 i\hbar a_\i} \prod_{j=1}^N 
  \bigg(\frac{\p \g_j}{\p w_j}\bigg)^{\f12 i\hbar a_j} \tP  \la{KZ1}
\ee
e.g.~for $SL(2,\R)$ we have $a_j=s_j(s_j-2)$ where $s_j\in
\{2,3,4,\dots\}$ is the (non-compact) spin of the $j$-th
representation. Note the striking similarity of the formula
\Ref{KZ1} with its classical analog \Ref{tauconf}, which was
already stressed at the beginning of this section.

Equation \Ref{KZ1} thus expresses the physical state $\Phi$ solving
\Ref{physstate2} as a product of an explicitly computable function $F$
and a solution of the KZ equation. There is a large body of literature
on the KZ equations, and although most of this work is concerned with
compact groups, explicit solutions based on the discrete
representations of $SL(2,\R)$ are known \ci{SchVar90,Babu93}. It would
thus appear that we simply have to insert these solutions into
\Ref{KZ1}, and we would be done.  However, we still have to take into
account the coset constraints \Ref{qcoset4}: as it turns out,
unfortunately, the solutions given above \Ref{KZ1} are not compatible
with these extra conditions. In fact, because the constraints are
second class the whole construction must be modified. Actually, the
solutions of \ci{SchVar90,Babu93} are already incompatible with the
single constraint $B_\i\!=\!0$, as can be readily seen: to satisfy
$B_\i=0$ or, equivalently, \Ref{qcoset5} (and thereby get rid of the
first factor in \Ref{KZ1}), we need $a_\i=0$; however, for the lowest
weight unitary representations of $SL(2,\R)$ used in \ci{SchVar90,
Babu93}, we have $a_\i = M + \sum_j s_j(s_j-2)$ where $M$ is a
positive integer, and the constraint can never be satisfied because
$s_j(s_j-2)\geq 0$ (using highest weight representations instead just
``inverts" the problem).  The constraint $B_\i=0$ could conceivably be
satisfied with continuous representations, or alternatively, by
simultaneous use of positive and negative (i.e.~both highest {\em and}
lowest weight) representations of $SL(2,\R)$, but no solutions of the
KZ equations of this type are presently known.\footnote{Let us
mention, that for {\em compact} coset spaces such as $SU(2)/U(1)$ the
constraint \Ref{qcoset5} can also be satisfied \ci{KorNic95b}.}

To really solve the complete set of coset constraints, we must explicitly
take into account the relations \Ref{qcoset4} (remember that we are
only discussing the case of real poles and $B_j\in{\mathfrak g}$).
This means that we express everything in terms of half of the variables.  
The quantum constraints \Ref{physstate2} are modified accordingly.
We just state the result, whose derivation is analogous to \Ref{KZ1}: 
if $\tP$ is a solution to the following modified (Coset)-KZ system
\be\la{KZcos}
\frac{\p\tP}{\p\g_j} = 
\left\{\sum_{k=1,k\not=j}^n\frac{1+\g_k/\g_j}
{\g_j-\g_k}\,\O_{jk} + 
\sum_{k=1}^n\frac{\g_k+1/\g_j}
{\g_j\g_k-1}\,\tilde{\O}_{jk}\right\}\tP
\ee
with 
\ba
\Omega_{jk} &=&  \frac1{i\hbar}{\rm tr} \, B_j B_k
~=~ i\hbar\big( \ft12 h_j \otimes h_k 
+ e_j \otimes f_k + f_j \otimes e_k \big) \\
\tilde{\O}_{jk} &=& \frac1{i\hbar}{\rm tr} \, \n(B_j) B_k ~=~ 
-i\hbar\big( \ft12 h_j \otimes h_k 
+ e_j \otimes e_k + f_j \otimes f_k \big), \la{Om2}
\ea
then the quantum constraints \Ref{physstate2} are solved by
\be
\Phi (x) = \prod_{j=1}^n 
  \bigg(\g_j^{-1}\frac{\p\g_j}{\p w_j}\bigg)^{i\hbar a_j} \tP  \la{KZ2}
\ee
The analogous formula was obtained in the classical theory
\Ref{tauconfcos}. One obvious technical difficulty with \Ref{Om2} is
that the operator $\tilde{\O}_{jk}$ does not preserve the excitation
number unlike \Ref{Om1}. This indicates that an ansatz of the type
used in \ci{SchVar90, Babu93} which starts from a state of fixed
occupation number will no longer work, so entirely new techniques may
be necessary to make progress with the above equations.  We would also
like to mention that for the ``really interesting" higher rank
non-compact groups (such as e.g.~$E_{8(+8)}$ for $N\!=\!16$
supergravity), the representation theory is much less developed
\ci{GunSac82}; for instance, although it is known that these groups do
admit discrete unitary irreducible representations which are neither
of highest nor lowest weight type, so far only preliminary results on
their explicit form are available.

Finally, our results show that, in the reduction to two dimensions,
the world-sheet itself has become a secondary object, while the
complex spectral parameter $\g$ emerges as the truly fundamental
variable. In this sense the theory has become effectively one-dimensional. 
In fact, as recently shown in \ci{JulNic96}, the theory is not only
generally covariant as a $2d$ world-sheet theory, but in addition
admits a kind of general covariance  w.r.t.~to the spectral 
parameters $\g$ and $w$ as well. Its inherent quantum group structure 
has been mentioned several times in these lectures. At a more 
technical level it was discussed in section 4.4, where the 
quantization of the quadratic algebra of observables \Ref{monoMiMj},
\Ref{monoMiMi} was shown to naturally imply a quantum group structure.
Perhaps the emergence of a ``quantum space time" (in the target space)
might find a natural explanation in these theories.

\section*{ACKNOWLEDGMENTS} H.~N. is grateful to the organizers for the
invitation to lecture at this Carg\`ese Summer School, and to the
participants for many stimulating discussions. He would also like to
thank A.~Ashtekar and M.~G\"unaydin for valuable comments.  The work
of D.~K. was supported by DFG Contract Ni~290/5-1; H.~S. thanks
Studienstiftung des Deutschen Volkes for support.

\begin{appendix}
\renewcommand{\thesection}{A}
\renewcommand{\theequation}{A.\arabic{equation}}
\setcounter{equation}{0}

\section*{APPENDIX: EXTENSION TO SUPERSYMMETRY}
In this appendix we briefly describe the extension of the results
given in the main body of these lectures to locally supersymmetric
models, i.e.~matter-coupled supergravities in two dimensions.  In
addition to the bosonic fields introduced in section~1, these models
contain matter fermions as well as $N$ gravitinos and dilatinos, which
are the superpartners of the zweibein and the dilaton, where $N\leq
16$ is the number of local supersymmetries.  The structure of the
relevant supermultiplets has been discussed in
\ci{Nico87,NicWar89,Nico94}, to which we refer the reader for further
details. As shown there, the equations of motion and the associated
linear system are considerably more complicated than for the bosonic
theory. Our account will therefore be quite sketchy.  To make life as
simple as possible, we only consider the linear system of \ci{Nico87}
where all terms containing gravitinos and dilatinos have been
eliminated. The linear system can then be cast into a form analogous
to \Ref{lslc}:
\be
\Psi^{-1}\p_{\pm}\Psi = 
\Big\{\f{2}{1\pm\g}\Ph_{\pm}+\f{\g}{(1\pm\g)^2} \Rh_{\pm}\Big\} 
\la{sls}\ee
where (cf.~\Ref{geqm})
\be
\Ph_\pm \equiv \n(\cV) P_\pm \n (\cV^{-1}) \;\;\; , \;\;\;
\Rh_\pm \equiv \n(\cV) R_\pm \n (\cV^{-1}) 
\ee
with $P_\pm\equiv P_\pm^A Z_A \in {\mathfrak k}$ defined as before 
(cf.~\Ref{PQ}), while $R_\pm \equiv R_\pm^\a Z_\a\in {\mathfrak h}$ 
is the following expression bilinear in the matter fermions 
\be
R_\pm := \bar \chi \gamma_\pm \Gamma(Z^\a) \chi \cdot Z_\a
           \in {\mathfrak h}
\ee
Here $\gamma_\pm$ are the standard $2d$ $\gamma$-matrices, and
the matter fermions $\chi$ belong to a spinorial representation of 
the gauge group $H$ which is generated by the matrices $\Gamma (Z^\a)$.

The presence of higher order poles in \Ref{sls} necessitates 
a modification of the isomonodromic ansatz \Ref{spcurrent}:
in addition to the ``movable" poles at $\g=\g_j(x)$ in \Ref{spcurrent} 
we must now allow for ``rigid" poles at the branch 
points $\g\!=\!\pm 1$. The modified isomonodromic ansatz is given by
\be
A(\g) := \Psi^{-1} \p_\g \Psi =
\sum_{j=1}^{N}\f{A_j}{\g-\g_j}+\f{A_+}{\g+1}+\f{A_-}{\g-1}
\la{isomon1}\ee
The residues will again be subject to certain reality and coset
constraints which are implied by $P_\pm \in {\mathfrak k}$ and 
$R_\pm \in {\mathfrak h}$. We will not discuss these constraints
here, save for remarking that the constraint
\be
A_\i \equiv \sum_{j=1}^N A_j + A_- + A_+ =0
\la{Bis}\ee 
will be understood to hold.

Analyzing the singular terms at $\g=\pm 1$ a calculation completely 
analogous to the one leading to \Ref{rel} now shows that we can 
again recover $\Rh_{\pm}$ and $\Ph_{\pm}$ in terms of the residues 
$A_j$ and $A_\pm$; the result is
\ba
\Rh_\pm &=& \pm 4 \r^{-1}\p_\pm\r A_\pm    \non
\Ph_\pm &=&  2 \r^{-1} \p_\pm\r  \left\{\pm \ft12 (A_+ + A_-) + 
\sum_{j=1}^N \f{A_j}{1\pm\g_j}\right\} 
\ea  

The compatibility conditions of the linear system \Ref{sls}
with the supersymmetrized isomonodromic ansatz \ref{isomon1} 
imply the following set of deformation equations:
\ba
\p_\pm A_j &=&  \r^{-1}\p_\pm\r 
\left\{2\sum_{k=1}^N \f{[A_k, A_j]}{(1\pm\g_k)(1\pm\g_j)}+
\f{1}{1\pm\g_j}[ A_\mp,\; A_j] + \f{1\pm 3\g_j}{(1\pm\g_j)^2}
[ A_\pm,\; A_j]\right\}   \non
\p_\pm  A_{\pm} &=&  \r^{-1} \p_\pm\r 
\left\{\sum_j\f{1\pm 3\g_j}{(1\pm\g_j)^2}[ A_j,\; A_\pm]
\pm [ A_-,  A_+]\right\} \non
\p_\pm A_{\mp} &=&  \r^{-1} \p_\pm\r   
\left\{\sum_j\f{1}{(1\pm\g_j)^2}[ A_j,\; A_\mp]
\pm [ A_+, A_-]\right\} 
\ea

The Poisson structure can also be generalized. As in section 4.2
it is most conveniently written down in terms of
\be
B(\g):= \n (\cV^{-1}) A(\g) \n (\cV)
\ee
where $A(\g)$ is given by \Ref{isomon1}. It turns out that 
bosonic brackets between $\Bs_j$ and $\Bs_k$ \Ref{resbracket} 
remain unaltered, while for the new variables $\Bs_\pm$ we have
to demand
\ben
\{\Bs^a_\pm, \Bs^b_\pm\}= {f^{ab}}_c \Bs^c_\pm
\een
Given this Poisson structure, the matter Hamiltonians $H_\pm$ 
governing the evolution in the $x^\pm$-directions are 
\ba 
H_\pm = \r^{-1} \p_\pm\r
&\bigg\{ &\sum_{k, j}\f{\tr \Bs_j \Bs_k}{(1\pm\g_j)(1\pm\g_k)} \non &&
+\sum_j \Big[\f{1}{1\pm\g_j} \tr \Bs_j \Bs_\mp +\f{1\pm3\g_j}{(1\pm\g_j)^2}
\tr \Bs_j \Bs_\pm\Big] +\tr \Bs_+ \Bs_-\bigg\}
\la{Hx}
\ea

The above results indicate that an exact quantization of the supersymmetric
models may be possible. Generally one would expect that the
best way to go about this task is to solve the supersymmetry constraints,
i.e.~the ``square roots" of the bosonic constraints, but this may 
no longer be true if supersymmetry is ``bosonized" as in \Ref{sls}. 
As usual, the first step of quantization would be the replacement of 
the above brackets by quantum commutators, such that in addition to the 
commutation relations \Ref{CR20} we get two more relations
\be
[\Bs^a_\pm, \Bs^b_\pm]=i\hbar {f^{ab}}_c \Bs^c_\pm
\ee
The Hilbert space corresponding to a fixed isomonodromic sector
would consequently contain two extra factors, 
\be
\cH=\cH_1\otimes...\otimes\cH_N\otimes \cH_+\otimes \cH_-
\ee
where as before $\cH_j$ and $\cH_\pm$ are unitary representation 
spaces of ${\mathfrak g}$ or ${\mathfrak g}_\C$. The bosonic quantum 
constraints are formally the same as in the bosonic theory,
but with the new Hamiltonians \Ref{Hx}.

\end{appendix}

\numbibliography

\bibitem{KorNic95a}D.~Korotkin and H.~Nicolai, Separation of variables
  and {H}amiltonian formulation for the {E}rnst equation, {\em
  Phys.~Rev.~Lett.} 74:1272 (1995).

\bibitem{KorNic95b}D.~Korotkin and H.~Nicolai, An integrable model of
  quantum gravity, {\em Phys. Lett.} B356:211 (1995).

\bibitem{KorNic96}D.~Korotkin and H.~Nicolai, Isomonodromic
  quantization of dimensionally reduced gravity, {\em Nucl.~Phys.}
  B475:397 (1996).

\bibitem{KorSam96}D.~Korotkin and H.~Samtleben, Quantization of coset
  space sigma models coupled to two-dimensional gravity, preprint
  DESY-96-130, hep-th/9607095, Hamburg (1996).

\bibitem{BJ1}B.~Julia, Group disintegrations, in: {\em Superspace
and Supergravity}, S.~Hawking and M.~Rocek (eds.), Cambridge
University Press, Cambridge (1980).

\bibitem{dila}T.~Banks and M.~O'Loughlin, Two-dimensional quantum
gravity in Minkowski space, {\em Nucl.~Phys.}  B362:649 (1991);\\
C.~Callan, S.~Giddings, J.~Harvey and A.~Strominger, Evanescent black
holes, {\em Phys.~Rev.} D45:R1005 (1992);\\
D.~Louis-Martinez, J.~Gegenberg and G.~Kunstatter, Exact Dirac
quantization of all 2-d dilaton gravity theories, {\em Phys.~Lett.}
B321:193 (1994);\\
A.~Strominger, Les Houches lectures on black holes, hep-th/9501071, in
{\em Les Houches Summer School, Session 62: Fluctuating Geometries in
Statistical Mechanics and Field Theory};\\
T.~Kl\"osch and T.~Strobl, Classical and quantum gravity in
(1+1)-dimensions, {\em Class.~Quant.~Grav.} 13:965; 13:2395 (1996);\\
D.~Cangemi, R.~Jackiw and B.~Zwiebach, Physical states in matter
coupled dilaton gravity, {\em Ann.~Phys.} 245:408 (1996);\\
M.~Cavagli\`a and V.~de Alfaro, Quantization of a 2-d minisuperspace
model in dilaton Einstein gravity, preprint DFTT-10-96, gr-qc/9602062
(1996);\\
A.T.~Filippov, Exact solutions of (1+1)-dimensional dilaton gravity
coupled to matter, {\em Mod. Phys.~Lett.} A11:1691 (1996).

\bibitem{Fadd84}L.~Faddeev, Integrable models in (1+1) dimensional
  quantum field theory, in: {\em Les Houches, Session XXXIX, Recent
  Advances in Field Theory and Statistical Mechanics}, J.-B.~Zuber and
  R.~Stora (eds.), North-Holland, Amsterdam (1984).

\bibitem{FadTak87}L.~Faddeev and L.~Takhtajan, {\em Hamiltonian
  Methods in the Theory of Solitons}, Springer-Verlag, Berlin (1987).

\bibitem{KniZam84}V.~Knizhnik and A.B.~Zamolodchikov, Current algebra
  and {W}ess-{Z}umino model in two-dimensi\-ons, {\em Nucl.~Phys.}
  B247:83 (1984).

\bibitem{Hawk84}S.W.~Hawking, Quantum cosmology, in: {\em Relativity,
  Groups and Topology}, B.~De Witt and R.~Stora (eds.), North Holland,
  Amsterdam (1984).

\bibitem{Ish91}C.J.~Isham, Conceptual and geometrical problems in
  quantum gravity, in: {\em Recent Aspects of Quantum Fields},
  H.~Mitter and H.~Gausterer (eds.), Springer Verlag, Berlin (1991).

\bibitem{Asht91}A.~Ashtekar, {\em Lectures on Nonperturbative
  Canonical Gravity}, World Scientific, Singapore, 1991; 
  Mathematical problems in nonperturbative general relativity, in:
  {\em Gravitation and Quantizations}, B.~Julia and J.~Zinn-Justin
  (eds.), North Holland, Amsterdam (1995).

\bibitem{BrMaGi88}P.~Breitenlohner, D.~Maison and G.~Gibbons,
Four-dimensional black holes from Kaluza-Klein theories, {\em
Comm.~Math.~Phys.} 120:295 (1988).

\bibitem{EinRos37}A.~Einstein and N.~Rosen, {\em J.~Franklin Inst.}
223:43 (1937).

\bibitem{KhanPen71}K.~Khan and R.~Penrose, Scattering of two impulsive
gravitational plane waves, {\em Nature} 229:185 (1971).

\bibitem{Nico91}H.~Nicolai, Two-dimensional gravities and
  supergravities as integrable systems, in: {\em Recent Aspects of
  Quantum Fields}, H.~Mitter and H.~Gausterer (eds.), Springer-Verlag,
  Berlin (1991).

\bibitem{Erns68}F.~Ernst, New formulation of the axially symmetric
  gravitational field problem, {\em Phys.~Rev.} 167:1175 (1968).

\bibitem{Kuch71}K.~Kucha\v{r}, Canonical quantization of cylindrical
  gravitational waves, {\em Phys.~Rev.} D 4:955 (1971).

\bibitem{AshPie96a}A.~Ashtekar and M.~Pierri, Probing quantum gravity
  through exactly soluble midi-superspaces {I}, preprint CGPG-96/5-3,
  gr-qc/9606085, Penn State (1996), J.~Math.~Phys.~in press.

\bibitem{BelZak78}V.~Belinskii and V.~Zakharov, Integration of the
  {E}instein equations by means of the inverse scattering problem
  technique and construction of exact soliton solutions, {\em
  Sov.~Phys.~JETP} 48:985 (1978).

\bibitem{Mais78}D.~Maison, Are the stationary, axially symmetric
  {E}instein equations completely integrable?, {\em Phys.~Rev.~Lett.}
  41:521 (1978).

\bibitem{BreMai87}P.~Breitenlohner and D.~Maison, On the {G}eroch
  group, {\em Ann.~Inst.~H.~Poincar{\'e}.~Phys.~Th{\'e}or.} 46:215
  (1987).

\bibitem{selfdual1}J.~Fletcher and N.M.J.~Woodhouse, Twistor
Characterization of stationary axisymmetric solutions of Einstein's
equations, in: {\em Twistors in Mathematics and Physics}, T.N.~Bailey
and R.J.~Baston (eds.), LMS Lecture Notes, Series 156, Cambridge
University Press, Cambridge (1990);\\
L.J.~Mason and N.M.J.~Woodhouse, The Geroch group and non-Hausdorff
twistor spaces, {\em Nonlinearity} 1:73 (1988).

\bibitem{selfdual2}F.A.~Bais and R.~Sasaki, On the complete
integrability of the static axially symmetric gauge field equations
for an arbitrary group, {\em Nucl.~Phys.} B195:522 (1982); On the
algebraic structure of selfdual gauge fields and sigma models, {\em
Nucl.~Phys.} B227:75 (1983).

\bibitem{selfdual3}S.~Chakravarty, L.J.~Mason and E.T.~Newman,
B\"acklund transformations for the antiselfdual Yang-Mills equations,
{\em J.~Math.~Phys.}  29:1005 (1988).

\bibitem{BJ2}B.~Julia, Infinite Lie algebras in physics, 
in {\em Johns Hopkins Workshop on Current
Problems in Particle Physics: Unified theories and Beyond}, Johns
Hopkins University, Baltimore (1981).

\bibitem{KinChi78}W.~Kinnersley and D.M.~Chitre, Symmetries of the
  stationary {E}instein-{M}axwell field equations III, {\em
  J.~Math.~Phys.} 19:1926 (1978).

\bibitem{JulNic96}B.~Julia and H.~Nicolai, Conformal internal symmetry
  of 2d sigma-models coupled to gravity and a dilaton, preprint
  LPTENS-96-38, DESY 96-124, hep-th/9608082 (1996), to appear in
  Nucl.~Phys.~B.

\bibitem{Nico94}H.~Nicolai, New linear systems for 2-d Poincar\'e
supergravities, {\em Nucl.~Phys.} B414:299 (1994).

\bibitem{JiMiMoSa80}M.~Jimbo, T.~Miwa, Y.~M\^{o}ri and M.~Sato,
  Density matrix of an impenetrable {B}ose gas and the fifth
  {P}ainlev\'e transcendent, {\em Physica} 1D:80 (1980).

\bibitem{JiMiUe81}M.~Jimbo, T.~Miwa and K.~Ueno, Monodromy preserving
  deformation of linear ordinary differential equations with rational
  coefficients, {\em Physica} 2D:306 (1981).

\bibitem{Dira49}P.M.~Dirac, Forms of relativistic dynamics {\em
Rev.~Mod.~Phys.} 21:392 (1949).

\bibitem{Hus96}V.~Husain, Einstein's equations and the chiral model,
{\em Phys.~Rev.} D53:4327 (1996). 

\bibitem{Mena96}G.A.~Mena Marugan, Canonical
quantization of cylindrically symmetric models, {\em Phys.~Rev.}
D53:3156 (1996).

\bibitem{Isham94}H.D.~Zeh, {\em The Physical basis of the Direction of
Time}, Springer-Verlag, Berlin (1992);\\ C.J.~Isham, Prima facie
questions in quantum gravity, in: {\em Canonical Relativity: Classical
and Quantum}, J.~Ehlers and H.~Friedrich (eds.), Springer-Verlag,
Berlin (1994).

\bibitem{Kief93}C.~Kiefer, The semiclassical approximation to quantum
  gravity, in: {\em Canonical gravity: from classical to quantum},
  J.~Ehlers and H.~Friedrich (eds.), Springer-Verlag, Berlin (1994).

\bibitem{Dira67}P.M.~Dirac, {\em Lectures on quantum
mechanics}, Academic Press, New York (1967).

\bibitem{HenTei92}M.~Hennaux and C.~Teitelboim, {\em Quantization of
  Gauge Systems}, Princeton University Press, Princeton (1992).

\bibitem{KSHM80}D.~Kramer, H.~Stephani, E.~Herlt and M.~MacCallum,
  {\em Exact Solutions of Einstein's Field Equations}, Cambridge
  University Press, Cambridge (1980).

\bibitem{KorNic94}D.~Korotkin and H.~Nicolai, The {E}rnst equation on
  a {R}iemann surface, {\em Nucl.~Phys.} B429:229 (1994).

\bibitem{Gowd74}R.H. Gowdy, Vacuum spacetimes with two-parameter
spacelike isometry group and compact invariant hypersurfaces:
topologies and boundary conditions, {\em Ann.~of Phys.} 83:203 (1974).

\bibitem{VeEiMa84}H.~de~Vega, H.~Eichenherr and J.~Maillet, Classical
  and quantum algebras of non-local charges in $\s$ models, {\em
  Commun.~Math.~Phys.} 92:507 (1984).

\bibitem{AlekC96}A.Y.~Alekseev, Lectures at this Summer School.

\bibitem{FocRos92}V.~Fock and A.~Rosly, Poisson structures on moduli
  of flat connections on {R}iemann surfaces and $r$-matrices, preprint
  ITEP 72-92, Moscow (1992).

\bibitem{AlGrSc95a}A.Y.~Alekseev, H.~Grosse and V.~Schomerus,
  Combinatorial quantization of the {H}amiltonian {C}hern-{S}imons
  theory, {\em Commun.~Math.~Phys.} 172:317 (1995).

\bibitem{Murat}W.~Schmidt, in {\em Representation Theory of Lie
Groups}, M.~Atiyah (ed.), Cambridge University Press, Cambridge
(1979); \\ S.~Lang, $SL_2(\R)$, Graduate Texts in Mathematics,
Springer Verlag, Berlin (1985).

\bibitem{GunSac82}M.~G\"unaydin and C.~Sa\c{c}lio\~glu, Oscillator like
  unitary representations of noncompact groups with a {J}ordan
  structure and the noncompact groups of supergravity, {\em
  Commun.~Math.~Phys.} 87:159 (1982).

\bibitem{SchVar90}V.V.~Schechtman and A.N.~Varchenko, Hypergeometric
  solutions of {K}nizhnik-{Z}amolodchikov equations, {\em
  Lett.~Math.~Phys.} 20:279 (1990).

\bibitem{Babu93}H.~Babujian, Off-shell Bethe ansatz equations and
$N$-point correlators in the $SU(2)$ WZNW theory, {\em J.~Phys.}
A26:6981 (1993).

\bibitem{Nico87}H.~Nicolai, The integrability of {$N$}=16
  supergravity, {\em Phys.~Lett.}  B194:402 (1996).

\bibitem{NicWar89}H.~Nicolai and N.~P.~Warner, The structure of
  {$N$}=16 supergravity in two dimensions, {\em Commun.~Math.~Phys.}
  125:369 (1989).

\endnumbibliography

\end{document}